\documentclass[12pt,a4paper]{article}

\def\ben{\begin{enumerate}} \def\een{\end{enumerate}}
\def\beq{\begin{equation}} \def\eeq{\end{equation}}
\def\beqn{\begin{equation*}} \def\eeqn{\end{equation*}}
\def\bea{\begin{eqnarray}} \def\eea{\end{eqnarray}}
\def\ba{\begin{array}} \def\ea{\end{array}}
\def\beann{\begin{eqnarray*}} \def\eeann{\end{eqnarray*}}
\def\beasn{\begin{sneqnarray}} \def\eeasn{\end{sneqnarray}}
\def\bi{\begin{itemize}} \def\ei{\end{itemize}}
\def\be{\begin{enumerate}} \def\ee{\end{enumerate}}

\usepackage{graphicx}
\usepackage{color}
\usepackage[paper=a4paper,left=30mm,right=30mm,top=30mm,bottom=30mm]{geometry}
\usepackage{amsmath}
\usepackage{amsthm}
\usepackage{amssymb}
\usepackage{enumerate}

\numberwithin{equation}{section}

\title{L\'eon Rosenfeld's general theory of constrained Hamiltonian dynamics}

\author
{Donald Salisbury,$^{1,2}$ and Kurt Sundermeyer,$^{1,3}$\\
\\
\normalsize{$^{1}$Max Planck Insitute for the History of Science,}\\
\normalsize{Boltzmannstrasse 22, 14195 Berlin, Germany}\\
\normalsize{$^{2}$Austin College, 900 North Grand Ave, Sherman, Texas 75090, USA}\\
\normalsize{$^{3}$Freie Universit\"at Berlin, Fachbereich Physik,}\\
\normalsize{ Berlin, Germany}\\
}

\begin{document} 


\baselineskip24pt


\maketitle


\begin{abstract}
This commentary reflects on the 1930 general theory of L\'eon Rosenfeld dealing with phase-space constraints. We start with a short biography of Rosenfeld and his motivation for this article in the context of ideas pursued by W. Pauli, F. Klein, E. Noether. We then comment on Rosenfeld's General Theory dealing with symmetries and constraints, symmetry generators, conservation laws and the construction of a Hamiltonian in the case of phase-space constraints. It is remarkable that he was able to derive expressions for all phase space symmetry generators without making explicit reference to the generator of time evolution. In his Applications, Rosenfeld  treated the general relativistic example of Einstein-Maxwell-Dirac theory. We show, that although Rosenfeld refrained from fully applying his general findings to this example, he could have obtained the Hamiltonian. Many of Rosenfeld's discoveries were re-developed or re-discovered by others two decades later, yet as we show there remain additional firsts that are still not recognized in the community. 
\end{abstract}


\section{Introduction}

L\'eon Rosenfeld's 1930 {\it Annalen der Physik} \cite{Rosenfeld:1930aa}\footnote{An English translation by D. Salisbury and K. Sundermeyer will appear in {\it European Physical Journal - H}.} paper developed a comprehensive Hamiltonian theory to deal with local symmetries that arise in Lagrangian field theory. Indeed, to a surprising degree he established the foundational principles that would later be rediscovered and in some respects extended by the individuals who until recently have been recognized as the inventors of the methods of constrained Hamiltonian dynamics, Peter Bergmann and Paul Dirac. Not only did he provide the tools to deal with the only local gauge symmetries that were known at the time, namely local $U(1)$ and local $Lorentz$ covariance, but he also established the technique for translating into a Hamiltonian description the general covariance under arbitrary spacetime coordinate transformations of Einstein's general theory of relativity. Some of this pioneering work either became known or was independently rediscovered over two decades later. But for unknown reasons Rosenfeld never claimed ownership, nor did he join later efforts to exploit his techniques in pursuing canonical approaches to quantum gravity.\footnote{He did present an unpublished seminar entitled ``Conservation theorems and invariance properties of the Lagrangian" in Dublin in May of 1946 where he repeated the invariance arguments but did not relate the discussion to phase space. Niels Bohr Archive, Rosenfeld Papers.}

He was brought to Zurich in 1929 by Wolfgang Pauli with the express purpose of helping to justify procedures that had been employed by Heisenberg and Pauli in their groundbreaking papers on quantum electrodynamical field theory. With the understanding that second quantization should naturally include all known fundamental interactions, Rosenfeld and Pauli apparently jointly decided that a new procedure was needed that would also take into account the dynamics of Einstein's gravitational field in interaction with electromagnetism and charged spinorial source fields. 

Among Rosenfeld's achievements are the following: He was the first to (1) Show that primary phase space constraints always arise as a consequence of local Lagrangian symmetries, (2) Show that local symmetries always involve singular Lagrangians, (3)  Exploit the identities that result from the symmetry transformation properties of the Lagrangian\footnote{These identities were first exploited by Felix Klein in the context of general relativity, as we shall discuss below.} to construct the constrained Hamiltonian that contains arbitrary spacetime functions, (4) Translate the vanishing conserved charge that arises as a consequence of symmetry transformations of the Lagrangian into a phase space expression, (5) Show explicitly that this symmetry generator, which we call the Rosenfeld-Noether generator, generates the correct infinitesimal transformations of all of the phase space variables, (6) Derive secondary and higher constraints through the requirement that primary constraints be preserved in time, (7) Show how to construct the constrained Hamiltonian and general covariance generator for general relativity - both for vacuum relativity and gravitation in dynamical interaction with the electromagnetic field and charged spinorial sources. Most of the advances listed here a now accepted wisdom - yet none have until recently been attributed to Rosenfeld.

Following a brief introduction to Rosenfeld in section 2 we will illustrate the seven accomplishments using two familiar simple models, the free electromagnetic field and the relativistic free particle. Then in section 4 we will present a detailed analysis of the first six of these achievements, referring to the general theory in Part 1 of his article. Where possible we employ Rosenfeld's notation. Section 5 is devoted to a description of the seventh achievement as it is related to Rosenfeld's general relativistic application. In section 6 we will take Rosenfeld's general findings and apply them to his example. Here we revert to modern notation and construct in detail the Hamiltonian and symmetry generators for Rosenfeld's tetrad gravity in interaction with the electromagnetic and spinorial fields. It is curious that he did not give the explicit expressions for the Hamiltonian in either the 1930 paper or the 1932 follow-up in which he reviewed the then current status of quantum electrodynamics.  In an Appendix we will give a capsule history of the later, better-known development of constrained Hamiltonian dynamics.

Before proceeding, it is clear from the title of Rosenfeld's article that he aimed at quantizing the Einstein-Maxwell-Dirac field. From the modern perspective he could perhaps be accused of a certain naivete in supposing that his fields could be promoted to quantum mechanical q-numbers through the simple expedient of forming a self-adjoint Hermitian operator by taking one half of the sum of the field operator and its Hermitian adjoint. But this is what he did in his equation preceding (R10). (Henceforth we will refer to his equations by adding the prefix R). The corresponding self-adjoint operators are expressed with an underline. This notation tends to make his text harder to read than necessary. And since, unless otherwise noted, we will be discussing the classical theory we will omit these underlines.

\section{Rosenfeld's personal background}

L\'eon Rosenfeld was born in 1904 in Charleroi, Belgium. After receiving his bachelor's  degree at the University of Li\`ege in 1926 he completed his graduate studies in Paris where under the supervision of Louis de Broglie and Th\'eophile de Donder he began his exploration of the link between quantum wave mechanics and general relativity\footnote{For more on Rosenfeld's life and collaborations see \cite{Jacobsen:2012aa}}. Thanks to the effort of de Donder at the 1927 Solvay Conference at Brussels, Rosenfeld secured a position as an assistant to Max Born in G\"ottingen. In G\"ottingen he found accomodations in the same home as Paul Dirac - who became a hiking partner. Emmy Noether was apparently temporarily in Russia during this time, and it is not clear whether he met her. Even if he had, given that her interests had shifted, it is unlikely that he would have discussed with her the second Noether theorem which plays a foundational role in this 1930 paper. 

This was a period of intense debate and evolving views regarding the recently established theories of wave and matrix mechanics, and Rosenfeld was ideally placed amidst the contenders. His position was somewhat unique given his working knowledge of general relativity and his previous efforts in unifying relativity with the incipient quantum wave theory. In addition to seeking an assistantship with Niels Bohr he also wrote to Albert Einstein, proposing that if he were successful with Einstein's aid in obtaining a research fellowship from the International Education Board he work under Einstein's supervision ``on the relations between quantum mechanics and relativity.'' \footnote{``Ich besch\"aftige mich mit den Beziehungen der Quantenmechanik zur Relativit\"atstheorie. Ihre Hilfe w\"are mir dabei von der gr\"ossten Wichtigkeit. Falls Sie damit einverstanden sind, dass ich unter Ihrer Leitung arbeite, bitte ich Sie um eine briefliche Mitteilung, die ich meinem Antrag beif\"ugen muss.'' Letter from Rosenfeld to Einstein, dated 26 April, 1928, Niels Bohr Archive, Rosenfeld Papers} Einstein replied almost immediately from Berlin, endorsing the project.\footnote{``Es freut mich, dass Sie \"uber den von Ihnen genannten Gegenstand im Zusammenhang mit mir Arbeiten wollen. Es w\"are gewiss erfreulich, wenn der International Education Board Ihnen zur Erm\"oglichung Ihres Aufenhaltes und Ihrer Arbeit in Berlin eine Fellowship gew\"ahren w\"urde.'', dated 3 May, 1928,   Niels Bohr Archive, Rosenfeld Papers} Rosenfeld also sought at the same time an arrangement with Niels Bohr who wrote back, advising according to Rosenfeld that coming to Copenhagen at the moment `` was not convenient and I had better postpone it". \footnote{Archive for the History of Quantum Physics (AHQP), 19 July, 1963, p. 5} Finally he wrote to Wolfgang Pauli who invited him to come to Zurich. Surprisingly, given Rosenfeld's general relativistic background, when asked by Pauli what he wished to do in Zurich, he replied that he intended to work on a problem involving the optical properties of metals. But he ``got provoked by Pauli to tackle this problem of the quantization of gravitation and the gravitation effects of light quanta."\footnote{AHQP, 19 July, 1963,p. 8} In an autobiographical note written in 1972 Rosenfeld says that in Zurich, where he arrived in the Spring of 1929, he ``participated in the elaboration of the theory of quantum electrodynamics just started by Pauli and Heisenberg, and he pursued these studies during the following decade; his main contributions being a general method of representation of quantized fields taking explicit account of the symmetry properties of these fields, a general method for constructing the energy-momentum tensor of any field, a discussion of the implications of quantization for the gravitational field $\ldots$'' \footnote{Niels Bohr Archive, Rosenfeld Papers}$\,$\footnote{For additional quantum electrodynamical background to Rosenfeld's 1930 paper see \cite{Salisbury:2009ab}}

\section{Two illustrative examples}

Before presenting our detailed discussion of Rosenfeld's general theory, we will illustrate its relevance with two familiar examples.  The first is the free electromagnetic field in flat spacetime. In the electromagnetic case the dynamical field is the vector potential  $A_\mu$ with associated field tensor $F_{\mu \nu} = A_{\nu, \mu} - A_{\mu, \nu}$ where $A_{\nu, \mu}:= \frac{\partial A_\nu}{\partial x^\mu}$. (We take the metric to have diagonal elements $(1,-1,-1,-1)$).
The flat space free electromagnetic field Lagrangian is
\beq
{\cal L}_{em} = -\frac{1}{4} F_{\mu \nu} F^{\mu \nu}.
\eeq
This Lagrangian is invariant under $U(1)$ gauge transformation with infinitesimal variations
\beq
\delta A_\mu = \xi_{,\mu}. \label{deltaem}
\eeq
The statement that the Lagrangian is invariant under the local symmetry (\ref{deltaem}) is the identity 
\beq
\delta {\cal L}_{em} =  \frac{\partial {\cal L}_{em}}{\partial A_{\mu,\nu}} \delta \left( A_{\mu,\nu}  \right) =  \frac{\partial {\cal L}_{em}}{\partial A_{\mu,\nu}}\xi_{,\mu \nu} \equiv 0. \label{emident}
\eeq
 (Note that $\frac{\partial {\cal L}_{em}}{\partial A_{\mu,\nu}} = F^{\mu \nu} = - F^{\nu \mu}$ and the identity results as a consequence of this anti-symmetry.) In particular, the coefficient of each distinct $\xi_{,\alpha \beta}$ vanishes identically when these coefficients are understood as functions of $A_{\mu,\nu}$. But now we introduce momenta $p^\alpha$ conjugate to the $A_\beta$. Defining $\dot A_\nu := A_{\nu,0}$, the momenta are defined to be $p^\alpha := \frac{\partial {\cal L}_{em}}{\partial \dot A_\alpha}$.
The seven Rosenfeld results applied to this model, numbered in parentheses, are 
\begin{enumerate}[{(1)}] 
\item There is a primary constraint expressing the vanishing of the coefficient of $\xi_{,00}$. It is $\frac{\partial {\cal L}_{em}}{\partial \dot A_0} = p^0 = 0$.
\item In making the transition to a Hamiltonian version of free electromagnetism we would like to be able to solve the defining equations for the momentum for the velocities $\dot A_\mu$ in terms of the $A_\nu$, $A_{\nu, a}$ ( where $ a$ is a spatial index), and $p^\alpha$. This is clearly not possible in this case since $\dot A_0$ does not even appear in these relations. Another way of viewing this problem is to note that the defining relations are linear in the velocities and take the form
$$
p^\alpha = \frac{\partial^2 {\cal L}_{em}}{\partial \dot A_\alpha \partial \dot A_\beta} \dot A_\beta + \frac{\partial^2 {\cal L}_{em}}{\partial \dot A_\alpha \partial  A_{\beta,b}} A_{\beta,b},
$$
then to solve for the velocities in terms of the momenta we would need to find the reciprocal of the Hessian matrix $ \frac{\partial^2 {\cal L}_{em}}{\partial \dot A_\alpha \partial \dot A_\beta}$. But this matrix is singular since $ \frac{\partial^2 {\cal L}_{em}}{\partial \dot A_\alpha \partial \dot A_0}= \frac{\partial p^0}{\partial \dot A_\alpha} = 0.$ It is singular as a consequence of the invariance of the Lagrangian under the gauge transformation (\ref{deltaem}).
\item  Since the time derivative of the nought component of the potential does not appear in the momenta, we can choose any value we wish for it without violating these relations. So let us take $\dot A_0 = \lambda$ where $ \lambda$ is an arbitrary spacetime dependent function. The remaining velocities can be solved, yielding $\dot A_a = p^a + V_{,a}$. Substituting into the canonical Hamiltonian we find
$$
{\cal H} = p^\alpha \dot A_\alpha - {\cal L}_{em}(A_{\mu,a}, \dot A_b(p^c,V_{,d})]
= \frac{1}{2} \left(p^a p^a +B_b B_b\right) + p^a A_{0,a} + \lambda p^0.
$$
The field $B_a = \epsilon_{abc} A_{b,c}$ is the magnetic field.
\item The identity (\ref{emident}) can be conveniently rewritten in terms of the Euler-Lagrange equations,
\beq
- \xi_{,\mu}\frac{\partial}{\partial x^\nu}\frac{\partial {\cal L}_{em}}{\partial A_{\mu,\nu}}+ \left(\frac{\partial {\cal L}_{em}}{\partial A_{\mu, \nu}}  \xi_{,\mu}\right)_{,\nu} \equiv 0.
\eeq
We deduce that when the Euler-Lagrange equations are satisfied we have a conserved charge
$$
\overline{{\cal M}}_{em} =:\int d^3x \left(p^0 \dot \xi - p^a_{,a} \xi  \right),
$$
where we have assumed that the arbitrary $\xi$ go to zero at spatial infinity. Since $\xi$ also has arbitrary time dependence it is clear that in addition to the primary constraint $p^0 = 0 $  we must also have a secondary constraint $p^a_{,a} = 0$.
\item The constraint $\overline{{\cal M}}_{em}$ generates the infinitesimal symmetry transformations
$$
\delta A_\mu = \left\{\delta A_\mu, \overline{{\cal M}}_{em}  \right\} = \xi_{,\mu},
$$
and $\delta p^a = 0$.
\item The deduction (4) may be understood as a derivation of a higher order(secondary) constraint in the sense that if we write $\overline{{\cal M}}_{em} = \int d^3x \left(p^0 \dot \xi + {\cal N} \xi  \right)$, then we have $\frac{d}{dt}\overline{{\cal M}}_{em} = \int d^3x \left(\dot p^0 \dot xi +p^0 \ddot \xi + \dot {\cal N} \xi + {\cal N} \dot \xi \right) = 0.$ The vanishing of the coefficient of $\dot \xi$ then yields $ \dot p^0 = - {\cal N} = 0$.
\item Since this model is not generally covariant the achievement number seven is not relevant.
\end{enumerate}

Our next model is generally covariant, and it will serve to display some important differences with models that obey internal gauge symmetries like the $U(1)$ symmetry. We consider the parameterized free relativistic particle. Let $x^\mu(\theta)$ represent the particle spacetime trajectory parameterized by $\theta$. Under a reparameterization $\theta' = f(\theta)$, where $f$ is an arbitrary positive definite function, $x^\mu$ transforms as a scalar,
$$
x'^\mu(\theta') = x^\mu(\theta).
$$
We introduce an auxiliary variable $N(\theta)$ and we assume that it transforms as a scalar density of weight one,
$$
N'(\theta') = N(\theta)\left|\frac{d\theta}{d\theta'}\right|^1.
$$
Then the particle Lagrangian takes the form
\beq
L_p(\dot x^\mu,N) = \frac{1}{2 N} \frac{dx^\mu}{d\theta} \frac{dx_\mu}{d\theta}- \frac{m^2 N}{2},
\eeq
where $\dot x^\mu := \frac{dx^\mu}{d\theta}$.
It is quadratic in the velocities and Rosenfeld's general theory is therefore directly applicable. 
The Lagrangian transforms as a scalar density of weight one under parameterizations, i.e., 
\beq
L'_p\left( \frac{dx'^\mu(\theta')}{d \theta'},N'(\theta')\right) = L_p\left( \frac{dx'^\mu(\theta')}{d \theta'},N'(\theta')\right) = L_p\left(\frac{dx^\mu(\theta)}{d \theta},N(\theta)\right) \frac{d\theta}{d\theta'}. \label{deltaL1}
\eeq
Consequently, the equations of motion are covariant under reparameterizations.

Now consider an infinitesimal reparameterization $\theta' = \theta + \xi(\theta)$ with corresponding variations
$$
\delta x^\mu(\theta) := x'^\mu\left(\theta + \xi(\theta)\right) -x^\mu\left(\theta\right) = 0,
$$
$$
\delta \left(\frac{dx^\mu}{d\theta}\right) := \frac{dx'^\mu(\theta')}{d\theta'}-\frac{dx^\mu(\theta)}{d\theta} = - \dot \xi(\theta) \dot x^\mu(\theta),
$$
and
$$
\delta N(\theta) := N'\left(\theta + \xi(\theta)\right) - N(\theta) = -N(\theta) \dot \xi.
$$
Then (\ref{deltaL1}) yields the identity
\beq
L_p \dot \xi + \frac{\partial L_p}{\partial \dot x^\mu} \delta (\dot x^\mu) + \frac{\partial L_p}{\partial N} \delta N \equiv 0. \label{partident2}
\eeq
Again it will be convenient to express this identity 
 in terms of the Euler-Lagrange equations. For this purpose we introduce the  $\delta^*$ variation associated with the infinitesimal reparameterization. (It is actually minus the Lie derivative.) To save writing we will represent the variables $x^\mu$ and $N$ by a generic $Q_\alpha$.
For an arbitrary function of variables $\Phi$ we define
$$
\delta^* \Phi(\theta) := \Phi'(\theta) - \Phi(\theta) = \delta \Phi(\theta) - \dot \Phi(\theta) \xi(\theta).
$$
This has the property that $\delta^*\left( \dot \Phi \right) = \frac{d}{d\theta} \left( \delta^* \Phi \right)$.

In terms of the $Q_\alpha$ the identity (\ref{partident2}) is 
\beq
L_p \dot \xi + \frac{\partial L_p}{\partial Q_\alpha} \delta Q_\alpha + \frac{\partial L}{\partial \dot Q_\alpha} \delta \left(\dot Q_\alpha\right) \equiv 0. \label{partident3}
\eeq
Note that
$$
\delta \left(\dot Q_\alpha\right) =  \frac{d}{d\theta} \delta Q_\alpha - \dot Q_\alpha \dot \xi.
$$
Thus we may rewrite (\ref{partident3}) as
\beq
\frac{\delta L_p}{\delta Q_\alpha} \delta Q_\alpha + \frac{d}{d\theta} \left(\frac{\partial L_p}{\partial \dot Q_\alpha}\delta Q_\alpha  \right) - \frac{\partial L_p}{\partial \dot Q_\alpha} \dot Q_\alpha \dot \xi + L_p \dot \xi \equiv 0, \label{partident4}
\eeq
where $\frac{\delta L_p}{\delta Q_\alpha} = \frac{\partial L_p}{\partial Q_\alpha} - \frac{d}{d\theta}\frac{\partial L_p}{\partial \dot Q_\alpha} = 0$ are the Euler-Lagrange equations. 

One final rewriting of this identity yields a conserved charge. Substituting $\delta Q_\alpha = \delta^* Q_\alpha + \dot Q_\alpha \xi$ we find
\beq
\frac{\delta L}{\delta Q_\alpha} \delta^* Q_\alpha + \frac{d}{d\theta} \left(\frac{\partial L}{\partial \dot Q} \delta Q_\alpha - \frac{\partial L}{\partial \dot Q} \dot Q_\alpha \xi + L \xi  \right) \equiv 0. \label{partident5}
\eeq

Proceeding with Rosenfeld's achievements applied to this model we have
\begin{enumerate}[{(1)}] 
\item The second derivative $\ddot \xi$ could arise in (\ref{partident2}) only if $\dot N$ were to appear in the Lagrangian, and this would spoil to reparameterization covariance. Thus we must have $p_N := \frac{\partial L_p}{\partial \dot N} = 0$.
\item The Hessian is singular since $\frac{\partial L_p}{\partial \dot N} = 0$.
\item We can take $\dot N = \lambda$ where $\lambda$ is positive-definite but otherwise an arbitrary function of $\theta$. The remaining velocities follow from the definitions
$$
p_\mu = \frac{\partial L_p}{\partial \dot x^\mu} = \frac{\dot x_\mu}{N}. 
$$
Solving for $\dot x^\mu$ we have 
$$
\dot x^\mu = N p^\mu.
$$
Substituting these velocities into the Hamiltonian we have
$$
H_p(p^\mu,\lambda) = p_\mu \dot x^\mu(p) + p_N \dot N - L_p(\dot x(p)) = \frac{N}{2} \left(p^2 + m^2  \right) + \lambda p_N.
$$
\item According to (\ref{partident5}) the conserved charge associated with the free relativistic particle is
\bea
M_{p} &=& p_\mu \delta x^\mu + p_N \delta N - p_\mu \dot x^\mu \xi - p_N \dot N \xi+ L_p \xi \nonumber \\
&=& - p_N N \dot \xi - \frac{N \xi}{2}\left(p^2 + m^2  \right) - \lambda p_N \xi,
\eea
where we have used the same procedure described in item (3) to obtain a phase space function involving also the arbitrary function $\lambda$.
\item $M_p$ generates the correct infinitesimal reparameterization symmetry variations.
$$
\delta^* x^\mu = \left\{x^\mu, M_p  \right\} = - N p^\mu \xi = - \dot x^\mu \xi.
$$
In the last equality we used the equation of motion. This is the correct $\delta^*$ variation for a scalar. Also we have
\beq
\delta^* N =  \left\{N, M_p  \right\} = - N \dot \xi - \lambda \xi = - N \dot \xi - \dot N \xi,
\eeq
where again in the last equality we used the equation of motion. This is the correct $\delta^*$ variation of a scalar density.
\item We deduce that in addition to the primary constraint $p_N = 0$ we have a secondary constraint $p^2 + m^2 = 0$.

\item This is a generally covariant model, and as we shall see, the construction of the generator of infinitesimal diffeomorphisms does also apply to general relativity. It is significant, however, that the charge we have obtained only works for infinitesimal variations. As we shall discuss in detail later, this deficiency is related to the fact that we need to apply the equations of motion in order to obtain the correct variations.

\end{enumerate}

\section{Rosenfeld's original contributions in the general theory} \label{3}

Concerning the invention of constrained Hamiltonian dynamics there is little in the work of Bergmann \cite{Bergmann:1949aa} , Bergmann and Brunings \cite{Bergmann:1949ab}, Dirac \cite{Dirac:1950aa,Dirac:1951aa}, Bergmann, Penfield, Schiller, and Zatkis \cite{Bergmann:1950ab}, Anderson and Bergmann \cite{Anderson:1951aa}, Heller and Bergmann \cite{Heller:1951ab}, and Bergmann and Schiller \cite{Bergmann:1953aa} that was not already achieved or at least anticipated by L\'eon Rosenfeld over twenty years earlier. He also pioneered the field of phase space symmetry generators.

Rosenfeld assumed that the Lagrangian was quadratic in the field velocities, taking the form
\beq
2 {\cal L} = Q_{\alpha,\nu} {\cal A}^{\alpha \nu, \beta \mu}(Q) Q_{\beta, \mu} +2 Q_{\alpha,\nu} {\cal B}^{\alpha \nu} (Q) + {\cal C}(Q), \label{quadL}
\eeq
in his equation (R1). The $Q$ represent arbitrary fields that can have components represented by the generic index $\alpha, \beta, $ etc. from the beginning of the Greek alphabet. The  $,\mu$ represents a derivative with respect to the spacetime coordinate.
The Lagrangians considered later by Bergmann, Dirac, and also Arnowitt, Deser, and Misner \cite{Arnowitt:1962aa} are of this form. He contemplated both general coordinate and local gauge transformations. In his General discussion Rosenfeld uses a latin index for all of these cases. Anticipating his later example we distinguish between descriptors of general coordinate transformations using a Greek index,
\beq
\delta x^\mu = \xi^\mu,  \label{xtrans}
\eeq
$U(1)$ transformations with no index $\xi$, and local Lorentz transformations with a latin index $\xi^r$. Rosenfeld does not make this distinction in his abstract formalism, and it is our hope that this notation will make his article more accessible. 

Accordingly, the symmetry variations of the field variables are 
\beq
\delta Q_\alpha(x) = c_{\alpha r} (x,Q) \xi^r (x)  + c^\sigma_\alpha (x,Q) \frac{\partial \xi }{\partial x^\sigma} +c^\sigma_{\alpha \mu} (x,Q) \frac{\partial \xi^\mu }{\partial x^\sigma}. \label{deltaQ}
\eeq
(Rosenfeld actually considered more general variations. See (R2). We shall also represent the time component with $0$ rather than $4$, and with these restricted variations we avoid more complicated expressions like (R18c) where several $4$ upper indices appear.) 

Rosenfeld lets a ``prime'' represent the transformed variable, and
$$
\delta Q_\alpha := Q'_\alpha(x+\delta x) - Q_\alpha(x).
$$
Rosenfeld also introduced $\delta^*$ variations with the definition
\beq
\delta^* \Phi (x)= \delta \Phi (x)- \frac{\partial \Phi(x)}{\partial x^\nu} \delta x^\nu, \label{deltastarQ}
\eeq
where $\Phi$ is any functional of $x$ and $Q(x)$ and $\frac{\partial \Phi(x)}{\partial x^\nu} $ is the partial derivative with respect to the spacetime coordinate. The $\delta^*$ variations are minus the Lie derivative in the direction $\delta x^\nu$. Utiyama \cite{Utiyama:1947ab} in 1947 followed Rosenfeld's lead in employing the $\delta^*$ notation. E. Noether \cite{Noether:1918aa} in 1918 denoted these variations in the functional form by $ \bar \delta$.  P. G. Bergmann \cite{Bergmann:1949aa}, beginning in 1949,  continued Noether's use of the $ \bar \delta$ notation. These variations are now called ``active'' variations.

Rosenfeld's analysis is based on the known transformation properties of the Lagrangian density under the variations (\ref{deltaQ}). He considered  two cases that were relevant to his application. 

Rosenfeld's Case 1 assumes that the Lagrangian transforms as a scalar density of weight one under arbitrary spacetime coordinate transformations. As he notes in his equation  (R12), this is the statement that under the transformations (\ref{xtrans}) the variation of the Lagrangian is $\delta {\cal L} \equiv - {\cal L} \xi^\mu_{,\mu}$. This was true for his general relativistic model in which he coupled the gravitational field in tetrad form to electromagnetism and a charged spinorial field. This action is manifestly a scalar density even though it is not the Hilbert action and it is not an invariant under local Lorentz transformations as we shall see in Section \ref{4}. Rosenfeld's Case 2 incorporates the required transformation property under this internal gauge transformation. Rosenfeld showed how the identities that arise in both cases can be exploited to construct not only the Hamiltonian but also the phase space generators of infinitesimal coordinate and local gauge transformations.

We will write the fundamental identities (R12), (R13) and (R14) in a form that incorporates both Cases 1 and 2 of Rosenfeld. The extra term $\left(\delta {\cal K}^\mu\right)_{,\mu}$ results from the fact that under the local Lorentz transformations with descriptors $\xi^r$  the Lagrangian is not invariant. Indeed, in Rosenfeld's case 2 in which these variations occur, $\delta {\cal L} = -\delta \left({\cal K}^\mu_{,\mu}  \right) = -(\delta {\cal K}^\mu)_{,\mu}$. The net Lagrangian variation is then
\beq
\delta {\cal L} = \frac{\partial {\cal L}}{\partial Q_\alpha} \delta Q_\alpha +  \frac{\partial {\cal L}}{\partial Q_{\alpha, \mu}} \delta \left(Q_{\alpha, \mu} \right) \equiv - {\cal L} \xi^\mu_{,\mu} - \left(\delta {\cal K}^\mu\right)_{,\mu},\label{ident1a}
\eeq
or equivalently 
\beq
0  
\equiv\frac{\delta {\cal L}}{\delta Q_\alpha}\delta Q_\alpha + \left(\frac{\partial {\cal L}}{\partial Q_{\alpha, \nu}} \delta Q_\alpha\right)_{,\nu} - \frac{\partial {\cal L}}{\partial Q_{\alpha, \nu}}
Q_{\alpha, \mu} \xi^\mu_{, \nu} + {\cal L} \xi^\mu_{, \mu} + \left(\delta {\cal K}^\mu\right)_{,\mu}, \label{ident1b}
\eeq
where $\frac{\delta {\cal L}}{\delta Q_\alpha} - \left(\frac{\partial {\cal L}}{\partial Q_\alpha}\right)_{,\mu}=0 $ are the Euler-Lagrange equations and
as stated above ${\cal K}^\mu$ varies only under internal symmetries with descriptors $\xi^r$.  Rosenfeld assumed it be linear in derivatives of the field,
\beq
{\cal K}^\mu = f^{ \alpha \mu \rho}(Q) Q_{\alpha, \rho}.
\eeq
In fact, since only $\delta Q_\alpha(x) = c_{\alpha r} (x,Q) \xi^r (x)  $ comes into play in the variation of ${\cal K}^\mu$,  it follows since the identity (\ref{ident1a}) cannot depend on second derivatives of the $\xi^r$ that the variation of ${\cal K}^\mu_{,\mu}$ takes the form
\beq
\delta {\cal K}^\mu_{,\mu} = \left(r^{\alpha \mu} c_{\alpha r} \xi^r    \right)_{,\mu}=:\left({\cal I}^{ \mu}_{ r} \xi^r    \right)_{,\mu},
\eeq
where  according to (R73) 
\beq
r^{\alpha \mu} := -\frac{\partial f^{\alpha \mu \rho}}{\partial Q_\beta} Q_{\beta,\rho} + \frac{\partial f^{\beta \mu \rho}}{\partial Q_\alpha} Q_{\beta,\rho}.
\eeq
Thus we will work with the identity (\ref{ident1a}) in the form
\beq
\delta {\cal L} := \frac{\partial {\cal L}}{\partial Q_\alpha} \delta Q_\alpha +  \frac{\partial {\cal L}}{\partial Q_{\alpha, \mu}} \delta \left(Q_{\alpha, \mu} \right) \equiv - {\cal L} \xi^\mu_{,\mu} - \left({\cal I}^{ \mu}_{ r} \xi^r    \right)_{,\mu}. \label{ident1aa}
\eeq
This identity incorporates (R12), (R13) and (R75).

Since according to (\ref{deltastarQ})
$$
\delta \left(Q_{\alpha, \nu}  \right) = \left(\delta Q_\alpha  \right)_{,\nu} - Q_{\alpha,\mu} \xi^\mu_{,\nu},
$$
we can  equivalently write (\ref{ident1b}) in the form,
\beq
0 \equiv\frac{\delta {\cal L}}{\delta Q_\alpha}\delta Q_\alpha + \left(\frac{\partial {\cal L}}{\partial Q_{\alpha, \nu}} \delta Q_\alpha\right)_{,\nu} - \frac{\partial {\cal L}}{\partial Q_{\alpha, \nu}}
Q_{\alpha, \mu} \xi^\mu_{, \nu} + {\cal L} \xi^\mu_{, \mu} +\left({\cal I}^{ \mu}_{ r} \xi^r    \right)_{,\mu}. \label{ident1bb}
\eeq
This relation does not appear explicitly in Rosenfeld's work, but he then exploited
 the several identities that follow from these fundamental identities, namely the identical vanishing of the coefficient of each derivative of the arbitrary $\xi^\mu$, $\xi$, and $\xi^r$. He was not the first to deduce these identities. This discovery can be traced to Felix Klein \cite{Klein:1918aa}, and although Rosenfeld did not specifically identify Klein's procedure, he did cite one of his essential results, namely the appearance of the four field equations that did not involve accelerations when using Einstein's 1918  Lagrangian that was quadratic in the time derivatives of $g_{\mu \nu}$ \cite{Einstein:1918ae}.\footnote{See his remark preceding equation (R120).}  In any case, Rosenfeld was the first to project these relations to phase space. We think it likely that it was the Klein procedure that Rosenfeld refered to in his introduction when he noted that ``in the especially instructive example of gravitation theory, Professor Pauli helpfully indicated to me a new method that allows one to construct a Hamiltonian procedure in a definitely simpler and natural way when identities are present".  Pauli had exploited one of these identities in his Encyclopedia of the Mathematical Sciences contribution on relativity \cite{Pauli:1921aa}, and had cited Klein. One might be justified in interpreting this sentence as a recognition by Rosenfeld that Pauli had communicated to him the fundamental ideas of the general theory presented in this paper. We will comment on  this hypothesis in our concluding remarks. \footnote{In fact, Pauli derived the contracted Bianchi identities in the same manner that was later employed by Bergmann for generally covariant theories \cite{Bergmann:1949aa}. He performed an integration by parts of the identity, and then let the $\xi^\mu$ on the boundary vanish. Pauli did not offer a genuinely Klein inspired approach until his updated annotated relativity article appeared in 1958 \cite{Pauli:2000aa}. He shared this derivation first in a letter dated  9 October 1957, addressed to Charles Misner \cite{Meyenn:2001aa}. } Indeed, the series of volumes was Klein's creation, and Klein carefully read the article and offered constructive criticism.\footnote{See the discussion of the article in \cite{Enz:2002aa}}

\subsection*{(1) Primary constraints}

Substituting (\ref{deltaQ}) into the identity (\ref{ident1aa}) we find that the identically vanishing coefficients of $\xi^\mu_{,\rho \sigma}$ are
\beq
\frac{\partial {\cal L}}{\partial Q_{\alpha, (\rho}} c^{\sigma)}_{\alpha \mu}  \equiv 0, \label{identpmu}
\eeq
while the coefficients of $\xi_{,\rho \sigma}$ give
\beq
\frac{\partial {\cal L}}{\partial Q_{\alpha, (\rho}} c^{\sigma)}_{\alpha}  \equiv 0. \label{identp}
\eeq
With regard to the remaining transformations,  the coefficient of $\xi^r_{,\mu}$ gives
\beq
\frac{\partial {\cal L}}{\partial Q_{\alpha, \mu}} c_{\alpha r} + {\cal I}^\mu_r \equiv 0. \label{identpr}
\eeq

After introducing the momenta ${\cal P}^\alpha : = \frac{\partial {\cal L}}{\partial \dot Q_\alpha}$ Rosenfeld obtains  the phase space constraints (R18c)
\beq
{\cal P}^\alpha c^0_{\alpha \mu}=:{\cal F}_\mu= 0, \label{pcmu}
\eeq
\beq
{\cal P}^\alpha c^0_{\alpha}=:{\cal F}= 0, \label{pc}
\eeq
and
\beq
{\cal P}^\alpha c^0_{\alpha r} + {\cal I}^0_r=:{\cal F}'_{r} = 0. \label{pcr}
\eeq
This last relation corresponds to (R79).\footnote{Rosenfeld actually defines ${\cal F}:= {\cal P}^\alpha c^0_{\alpha r}$, in his Case 2. Thus in an effort to introduce a unified and hopefully more comprehensible notation, we are representing the actual constraint with a `prime'.} Looking at the vanishing coefficient of $Q_{\alpha, \mu \nu}$ in the identity (\ref{ident1bb}) under the variations $\delta Q_\alpha = c_{\alpha r} \xi^r$, Rosenfeld showed in (R80) that ${\cal I}^0_r$ is independent of $\dot Q_\alpha$. 
Thus the three relations (\ref{pcmu})-(\ref{pcr}) are primary constraints, using the terminology introduced by Anderson and Bergmann  in 1949 \cite{Anderson:1951aa}. 

\subsection*{(2) Singular Lagrangians}

For the quadratic Lagrangian the momenta take the form
\beq
{\cal P}^{\alpha } = {\cal A}^{\alpha  \beta } \dot Q_{\beta} + {\cal D}^{\alpha}, \label{palpha}
 \eeq
 where  ${\cal A}^{\alpha  \beta }$  and ${\cal D}$ are functions of $Q_\gamma$ and their spatial derivatives. Then  (\ref{identpmu}), (\ref{identp}) and (\ref{identpr}) deliver the additional identities,
 \beq
 c^0_{\alpha \mu} {\cal A}^{\alpha  \beta } = c^0_{\alpha} {\cal A}^{\alpha  \beta } = c^0_{\alpha  r} {\cal A}^{\alpha  \beta }\equiv 0, 
 \eeq
 and
 \beq
 c^0_{\alpha  \mu} {\cal D}^{\alpha} =c^0_{\alpha} {\cal D}^{\alpha} = c^0_{\alpha  r} {\cal D}^{\alpha} \equiv 0,
 \eeq
corresponding to (R25) and (R26). 
 The first is the statement that the $c^0_{\alpha \mu}$, $c^0_{\alpha}$ and $c^0_{\alpha  r}$ are null vectors of the Hessian matrix\footnote{Cecile DeWitt-Morette indicated to one of us several years ago that she denotes this the ``Legendre matrix", but the Hessian terminology now seems to be widespread.} ${\cal A}^{\alpha  \beta }$. As we shall see, Rosenfeld used all of these relations in his construction of the Hamiltonian.
 
\subsection*{(3) Construction of the Hamiltonian} 
 
In solving (\ref{palpha}) for the velocities, Rosenfeld refered to ``the theory of linear equations'' but did not give an explicit reference. His procedure was unique as far as we can tell. Here we repeat Rosenfeld's argument in \S 3, filling in some additional details to make the argument more comprehensible.

He first supposed that he had found, presumeably through a suitable linear combination of the linear equations (\ref{palpha}), a non-singular submatrix of the Hessian matrix of rank $N-r_0$, where $N$ is the total number of $Q_\alpha$ variables, and $r_0$ is the number of primary constraints. Label the indices of the non-singular matrix by $\alpha'$ and the remaining indices by $\alpha''$. Let ${\cal A}_{\alpha' \beta'}$  represent the inverse of the non-singular ${\cal A}^{\alpha' \beta'}$, i.e., ${\cal A}_{\alpha' \beta'} {\cal A}^{ \beta' \gamma'} = \delta^{\gamma'}_{\alpha'}$. Then the 
following $c^{\gamma''}_{r \alpha}$, where $\gamma'' = N-r_0+r$, are for each $\gamma''$ explicit null vectors of the matrix ${\cal A}^{\alpha \rho'}$,
$$
c^{\gamma''}_{r \alpha'} := {\cal A}^{\gamma'' \beta'}{\cal A}_{ \beta' \alpha'},
$$
and
$$
c^{\gamma''}_{r \alpha''} := - \delta^{\gamma''}_{\alpha''}.
$$
Explicitly, 
$$
c^{\gamma''}_{r \alpha} {\cal A}^{\alpha \rho'} = c^{\gamma''}_{r \alpha'} {\cal A}^{\alpha' \rho'}+c^{\gamma''}_{r \alpha''} {\cal A}^{\alpha'' \rho'} = 0.
$$
Now since these null vectors must be expressible as linear combinations of the $c_{\alpha\, r}^{0}$, $c_{\alpha\, \mu}^{0}$ and  $c_{\alpha}^{0}$, it follows that since $c_{\alpha\, \mu}^{0} {\cal P}^\alpha =c_{\alpha\, r}^{0} {\cal P}^\alpha = c_{\alpha}^{0} {\cal P}^\alpha\equiv 0$,
$$
0\equiv  {\cal A}^{\gamma'' \beta'}{\cal A}_{ \beta' \alpha'} {\cal P}^{\alpha'} - \delta^{\gamma''}_{\alpha''} {\cal P}^{\alpha''},
$$
and we have therefore solved for $ {\cal P}^{\alpha''}$ as a linear combination of the ${\cal P}^{\alpha'}$,
$$
{\cal P}^{\alpha''} \equiv {\cal A}^{\alpha'' \beta'}{\cal A}_{ \beta' \alpha'} {\cal P}^{\alpha'}.
$$
A similar relation holds for ${\cal D}^{\alpha''}$.
It follows that a special solution $\dot Q^0_\alpha$ of (\ref{palpha}) is (R32),
\beq
\left. \ba{l} \dot Q_{\beta'}^0 =  {\cal A}_{\beta' \gamma'} ( {\cal P}^{\gamma'} - {\cal D}^{\gamma'})  \\
   \dot Q_{\beta''}^0 = 0. \ea \right\} \label{32}
   \eeq
The general solution is therefore
 \beq
 \dot Q_\alpha = \dot Q_{\alpha}^0 + \lambda^\mu c^0_{\alpha \mu}+\lambda c^0_{\alpha }+\lambda^r c^0_{\alpha r}, \label{gensol}
 \eeq
 where the $\lambda^\mu$, $\lambda$ and $\lambda^r$ are arbitrary functions.
 
This method for solving linear singular equations is to be contrasted with a procedure pursued by Bergmann and his collaborators, beginning in 1950 \cite{Bergmann:1950aa}. His group employed the so-called ``quasi-inverses'', but again without explicit references. The procedure was first published by E. H. Moore in 1920 \cite{Moore:1920aa}. It was subsequently rediscovered and extended by R. Penrose in 1955 \cite{Penrose:1955aa}. Dirac invented his own idiosyncratic method in 1950 \cite{Dirac:1950aa}.\footnote{Dirac's point of departure was his assumption that the Hamiltonian $p^\alpha \dot Q_\alpha - L(Q, \dot Q)$ could be conceived as a function of independent variables $ Q$, $\dot Q$, and $p$.}

Rosenfeld then substituted the general solutions (\ref{gensol}) into the standard Hamiltonian 
\beq
{\cal H} = \dot Q_\alpha {\cal P}^\alpha - {\cal L}(Q, \dot Q).
\eeq
Explicitly, we have
$$
 {\cal L}(Q,  {\dot Q}{}^0 +   \lambda^\mu c^0_{\alpha \mu}+\lambda^r c^0_{\alpha r}+\lambda c^0_\alpha ) = \frac{1}{2} {\cal A}^{\alpha \beta} \dot Q{}^0_\alpha  \dot Q{}^0_\beta+ {\cal B}^{\alpha} \dot Q{}^0_\alpha + {\cal E}
$$
$$
=\frac{1}{2} {\cal A}_{\alpha' \beta'} {\cal P}^{\alpha'}  {\cal P}^{\beta'}+  {\cal A}_{\alpha' \beta'} {\cal B}^{\alpha'} {\cal P}^{\beta'} + {\cal E}
$$
where ${\cal E} := \frac{1}{2}{\cal A}^{\alpha a \beta b}  Q_{\alpha, a} Q_{\beta, b}+\frac{1}{2} {\cal C}$ and ${\cal C}$ is given in (\ref{quadL}).   Then since 
$$
{\cal P}^\alpha \dot Q_\alpha = {\cal A}_{\alpha' \beta'} {\cal P}^{\alpha'}  {\cal P}^{\beta'} +\lambda^\mu {\cal F}_\mu+\lambda {\cal F}+\lambda^r {\cal F}_r,
$$
we have finally Rosenfeld's (R35),
\beq
{\cal H} = {\cal P}^\alpha \dot Q_\alpha -  {\cal L} = {\cal H}_0 +\lambda^\mu {\cal F}_\mu+\lambda {\cal F}+ \lambda^r {\cal F}_r, \label{ham}
\eeq
where
$$
{\cal H}_0 = \frac{1}{2}{\cal A}_{\alpha' \beta'} {\cal P}^{\alpha'}  {\cal P}^{\beta'} -  {\cal A}_{\alpha' \beta'} {\cal B}^{\alpha'} {\cal P}^{\beta'} - {\cal E}.
$$

The Hamilton equations follow as usual from the variation of the Hamiltonian density,
$$
\delta {\cal H} = \delta {\cal P}^{\alpha} \dot Q_\alpha +{\cal P}^{\alpha} \delta \dot Q_\alpha - \left.\frac{ \delta L\left[Q, \dot Q[Q,{\cal P},\lambda]\right]}{\delta Q_\alpha}\right|_{\dot Q} \delta Q_\alpha - \frac{ \delta L\left[Q, \dot Q[Q,{\cal P},\lambda]\right]}{\delta \dot Q_\alpha} \delta \dot Q_\alpha
$$
$$
= \delta {\cal P}^{\alpha} \dot Q_\alpha  - \dot {\cal P^\alpha} \delta Q_\alpha,
$$
where we used (R33) and the Euler-Lagrange equations. Rosenfeld did not posit a new variational principle.\footnote{See \cite{Goldstein:1965aa}, for example, where in the context of non-singular systems one speaks of the `modified Hamilton's principle' $\delta \int_{t_1}^{t_2} \left(p^i \dot q_i - L(q,p,t)  \right) = 0$.} He simply proved that the Hamiltonian equations, with the Hamiltonian containing the arbitrary functions $\lambda^\mu$, $\lambda$ and $\lambda^r$, are equivalent to the Euler-Lagrange equations.

\subsection*{(4) Diffeomorphism and gauge generators}

For the purpose of constructing the phase space generators of infinitesimal symmetry transformations it is convenient to rewrite the  identity (\ref{ident1bb}) equivalently as
\beq
0 \equiv\frac{\delta {\cal L}}{\delta Q_\alpha}\delta^* Q_\alpha + \left[\frac{\partial {\cal L}}{\partial Q_{\alpha, \mu}}\delta Q_\alpha - \frac{\partial {\cal L}}{\partial Q_{\alpha, \mu}} Q_{\alpha, \nu} \xi^\nu + {\cal L} \xi^\mu+{\cal I}^\mu_r \xi^r  \right]_{,\mu}, \label{ident2bb}
\eeq
as Rosenfeld does explicitly in (R56c) for his Case 1. 

This form of the identity is actually the basis of Noether's second theorem. She gives the scalar density form explicitly in her equation (13) \cite{Noether:1918aa}.  Noether had also considered the case where the Lagrangian differed from a scalar density by a total divergence, referring to Einstein's quadratic Lagrangian and Klein's second note, his equation (30) \cite{Klein:1918aa}. This is the Lagrangian employed by Bergmann and his collaborators. Bergmann's student, Ralph Schiller, based his dissertation on the straightforward extension of Rosenfeld's technique to this case in which the divergence term is not invariant under coordinate transformations \cite{Schiller:1952aa}.\footnote{``With this information it is possible to show that the $\bar C$ of (6.8) is actually the generator of the $\bar \delta y_A$ and the $\bar \delta \pi^A$ transformations. The calculation is straightforward and closely follows a similar calculation in Rosenfeld,${}^{13}$ so that we shall not carry it out". The reference is to the paper we are analyzing here.}

As written our relation also incorporates Rosenfeld's Case 2. Indeed Case 2 deals with what  is often called quasi-invariance, namely invariance of a Lagrangian up to a total derivative. The inclusion of additional total derivative terms in Noether's theorems is traditionally attributed to Klein's assistant, Bessel-Hagen \cite{Bessel-Hagen:1921aa}. In Bessel-Hagen's own words ``First I give the two E. Noether theorems, actually in a somewhat more general form than they appear in the cited article. I owe this [new form] to a verbal communication from Fr\"aulein Emmy Noether herself". \footnote{``Zuerst gebe ich die beiden E. Noetherschen S\"atze an, und zwar in einer etwas allgemeineren Fassung als sie in der zitierten Note stehen. Ich verdanke diese einer m\"undlichen Mitteilung von Fr\"aulein Emmy Noether selbst."}  One might conclude that Noether was essentially involved in his work \cite{Kosmann-Schwarzbach:2011aa}. 

On-shell, that is on the solutions of the Euler-Lagrange equation $\frac{\delta {\cal L}}{\delta Q_\alpha}= \frac{\partial  {\cal L}}{\partial Q_\alpha} - \frac{\partial}{\partial x^\mu} \frac{\partial  {\cal L}}{\partial Q_{\alpha,\mu}} = 0$, the identity (\ref{ident2bb}) implies that the current 
$$
{\cal M}^\mu := \frac{\partial {\cal L}}{\partial Q_{\alpha, \mu}}\delta Q_\alpha - \frac{\partial {\cal L}}{\partial Q_{\alpha, \mu}} Q_{\alpha, \nu} \xi^\nu + {\cal L} \xi^\mu+{\cal I}^\mu_r \xi^r 
$$
is conserved, from which by applying the Gau{\ss} integral theorem one obtains a conserved charge
\bea
M[\xi] &:=&\int d^3\! x {\cal M}^0 = \int d^3\! x \big(\frac{\partial {\cal L}}{\partial Q_{\alpha, 0}}\delta Q_\alpha - \frac{\partial {\cal L}}{\partial Q_{\alpha, 0}} Q_{\alpha, \nu} \xi^\nu + {\cal L} \xi^0+{\cal I}^0_r  \xi^r\big) \nonumber \\
 &=& \int d^3\! x\left(\mathcal{P}^\alpha \delta Q_\alpha - {\cal H} \xi^0 - \mathcal{P}^\alpha Q_{\alpha,a} \xi^a + {\cal I}^0_r \xi^r  \right) \nonumber \\
 &=& \int d^3\! x\left(\mathcal{P}^\alpha \delta Q_\alpha - {\cal G}_\mu \xi^\mu  + {\cal I}^0_r \xi^r  \right) \label{M0}
\eea
where we introduced the momenta, the Hamiltonian density (\ref{ham}) and the energy-momentum density ${\cal G}_\mu = \mathcal{P}^\alpha Q_{\alpha,\mu}- \delta^0_\mu {\cal L}$, Rosenfeld's (R41). 

Rosenfeld gives no reference for these constructions, but it is most likely that he learned of these objects from Pauli \cite{Pauli:1921aa}, who in turn refers to F. Klein \cite{Klein:1918aa}. The pseudo-tensor was in fact first written down by Einstein \cite{Einstein:1918ae}, and that publication stimulated the symmetry analysis of E. Noether \cite{Noether:1918aa} and Klein. 

Rosenfeld was the first to promote the vanishing  charge (\ref{M0}) to a phase space symmetry generator, and also the first to show that it is a linear combination of  phase space constraints. Up to the time that Rosenfeld accomplished this feat, attention had been paid only to the nonvanishing conserved Noether charges that follow from global symmetries. 
Strangely, although it is manifestly evident in Rosenfeld's analysis, he never stated explicitly that this charge was constrained to vanish. 

\subsection*{(5) Infinitesimal variations generated by the Rosenfeld-Noether generator}

In a {\it tour de force} Rosenfeld proved that the charges (\ref{M0}) generated the correct $\delta^* Q_\alpha$ and $\delta^* p^\alpha$ variations of all of the canonical variables under all of the infinitesimal symmetry transformations.This is obvious for the configuration variables $Q_\alpha$ since 
\bea
\left\{Q_\alpha,\int d^3x {\cal M}^0\right\} 
&=& \left\{Q_\alpha, \int d^3x \left(p^\beta \delta Q_\beta - {\cal H} \xi^0 - {\cal G}_a \xi^a \right)  \right\} \nonumber \\
&=& \delta Q_\alpha - Q_{\alpha, \mu} \xi^\mu = \delta^* Q_\alpha.
\eea
It is less obvious for variations of the momenta, but Rosenfeld gives an explicit proof. Bergmann and his collaborators, who did over twenty years later consider the realization of general coordinate transformations as canonical transformations, did not provide an analogous proof. 

Rosenfeld showed in the equation preceding (R51) that for the generalized momenta ${\cal P}^{\alpha \nu}:= \frac{\partial {\cal L}}{\partial Q_{\alpha,\mu}}$,
\beq
\delta {\cal P}^{\alpha \nu} =  \frac{\partial \delta {\cal L}}{\partial Q_{\alpha, \nu}} - {\cal P}^{\beta \mu} \frac{ \partial \delta (Q_{\beta, \mu} )}{\partial Q_{\alpha , \nu}} = \frac{\partial \delta {\cal L}}{\partial Q_{\alpha, \nu}}  -{\cal P}^{\beta \nu} \frac{\partial \delta Q_\beta}{\partial Q_\alpha}  + {\cal P}^{\alpha \mu}\xi^\nu_{, \mu}. \label{deltap}
\eeq
Then he used the identity (\ref{ident1aa})  to conclude that
\bea
\delta {\cal P}^{\alpha \nu} =  -\xi^\mu_{,\mu}\frac{\partial \cal L}{\partial Q_{\alpha, \nu}} - \frac{\partial }{\partial Q_{\alpha, \nu}} \left({\cal I}^\mu_r \xi^r\right)_{,\mu} - {\cal P}^{\beta \mu} \left( \frac{\partial \delta Q_\beta}{\partial Q_\alpha} \delta^\nu_\mu - \xi^\nu_{, \mu} \delta^\alpha_\beta \right).
\eea
Therefore
\bea
\delta {\cal P}^\alpha &=&   -\xi^\mu_{,\mu}\frac{\partial \cal L}{ \partial \dot Q_{\alpha}} - \frac{\partial }{\partial \dot Q_{\alpha}} \left({\cal I}^\mu_r \xi^r\right)_{,\mu} - {\cal P}^{\beta \mu} \left( \frac{\partial \delta Q_\beta}{\partial Q_\alpha} \delta^\nu_\mu - \xi^\nu_{, \mu} \delta^\alpha_\beta \right) \nonumber \\
&=& -\xi^a_{,a}  {\cal P}^\alpha- \frac{\partial {\cal I}^0_r}{\partial  Q_{\alpha}}  \xi^r- {\cal P}^\beta \frac{\partial \delta Q_\beta}{\partial Q_\alpha} + {\cal P}^{\alpha a} \xi^0_{,a},
\eea
and thus
\beq
\delta^* {\cal P}^\alpha = -\xi^a_{,a}  {\cal P}^\alpha- \frac{\partial {\cal I}^0_r}{\partial  Q_{\alpha}}  \xi^r- {\cal P}^\beta \frac{\partial \delta Q_\beta}{\partial Q_\alpha} + {\cal P}^{\alpha a} \xi^0_{,a} -  {\cal P}^\alpha_{, \mu} \xi^\mu.
\eeq
This is indeed the variation generated by $\int d^3x {\cal M}^0$.

\subsection*{(6) Generation of secondary constraints}

Although Rosenfeld entitles his paragraph \S 7 as ``The infinitesimal transformations $\overline{\cal M}$ as integrals of the motion", we find in this section the derivation of what is called today secondary, tertiary $\ldots$ constraints. This remarkable procedure derives constraints without making explicit use of the Hamiltonian!

Let us rewrite the charge density in (\ref{M0}) using the transformations (\ref{deltaQ}) and taking into account the primary constraints (\ref{pcmu}), (\ref{pc}) and (\ref{pcr}). Identifying the coefficients of time derivatives of the descriptors as in (R59), we have
\bea
{\cal M}^0 &=& {\cal N}^1_\mu \dot \xi^\mu+{\cal N}^1_r \dot \xi^r +{\cal N}^1 \dot \xi +{\cal N}^0_\mu  \xi^\mu+{\cal N}^0  \xi +{\cal N}^0_r  \xi^r  \nonumber \\
&=& {\cal F}_\mu \dot \xi^\mu +{\cal F} \dot \xi + {\cal F}'_r \xi^r+ {\cal P}^\alpha c^a_{\alpha \mu} \xi^\mu_{,a} + {\cal P}^\alpha c^a_\alpha \xi_{,a}   -  {\cal H} \xi^0 - {\cal G}_a \xi^a \label{noether}
\eea

It is obvious from the conservation of Noether charge that follows from (\ref{ident1bb}) that the coefficients of each of the time derivatives of the arbitrary descriptors that appear in the charge density must vanish. Also, as Rosenfeld noted, the coefficients of the highest time derivative of the descriptor $\xi$ are constraints, now called primary constraints. But then he noted that setting equal to zero the time derivative of the charge density yielded a recursion relation among the vanishing coefficients. In particular, employing an integration by parts we obtain his relations (R63)
\beq
{\cal N}^1_0 = \dot {\cal F}_0 ={\cal N}^0_0 = \left({\cal P}^\alpha c^a_{\alpha 0}\right)_{, a} + {\cal H} = 0,
\eeq
\beq
{\cal N}^1_b =\dot {\cal F}_b ={\cal N}^0_b  = \left({\cal P}^\alpha c^a_{\alpha b}\right)_{, a} + {\cal G}_b = 0,
\eeq
\beq
{\cal N}^1 =\dot {\cal F} ={\cal N}^0  = \left({\cal P}^\alpha c^a_\alpha\right)_{, a}  = 0,
\eeq
In other words, he derived secondary constraints and in principle by (R63) tertiary and so on.

Already in 1930 Rosenfeld laid out a valid procedure for constructing the infinitesimal phase space generators of field variations produced by arbitrary infinitesimal coordinate transformations and also internal gauge transformations. The procedure is fully equivalent to the method employed by Bergmann and his collaborators in 1951. One remarkable and generally unrecognized feature of Rosenfeld's work is that he showed that preservation of primary constraints could lead to further constraints that he could construct explicitly. Algorithms for determining secondary and higher constraints have until now been attributed to Bergmann and collaborators, and also to Dirac. Regarding the diffeomorphism symmetry, Dirac never concerned himself, as did Rosenfeld and Bergmann, with the realization of this group as a phase space transformation group. 

In fact, it turns out that neither Rosenfeld, nor initially Bergmann, were able to implement finite diffeomorphism transformations. Rosenfeld implicitly acknowledged this failure (see his \S6) while Bergmann did observe that it was a crucial invention of Dirac that rendered possible the realization of a diffeomorphism-induced group. 

The problem with the Rosenfeld and the Bergmann-Anderson generators \cite{Anderson:1951aa} are two-fold. Under finite transformations the arbitrary functions $\lambda^r$ that appear in the Hamiltonian appear with time derivatives of infinite order - as do the coordinate transformation functions. The same is true for the generators that were rediscovered in 1982 by Castellani \cite{Castellani:1982aa}. We will address this point in section 4.4.

\section{Rosenfeld's application} \label{4}

We will now apply Rosenfeld's program to his Einstein-Maxwell-Dirac model, obtaining explicit expressions for the symmetry generators that appear in his 1930 article. In the final subsection we will apply Rosenfeld's general formalism to construct the Hamiltonian for the model. He did not display this expression, and if he did not actually derive it there is good reason to believe that he could have if he had so wished.

\subsection{The Einstein-Maxwell-Dirac theory}

\subsubsection*{The gravitational action}

We will translate Rosenfeld's notation into conventional contemporary form. He used Fock's conventions regarding the tetrads \cite{Fock:1929aa} and he employed a Minkowski metric with signature $-2$. We will denote Minkowski indices with capitalized latin letters from the middle of the alphabet, so the components of the Minkowski metric are
$$
\eta_{IJ} = \left( \ba{cccc}1&0&0&0 \\ 0&-1&0&0 \\0&0&-1&0 \\0&0&0&-1 \ea \right).
$$
Then his $h_{i,\nu}$ is the covariant tetrad with the Minkowski index lowered:  $h_{i,\nu} \hat = e_{I \nu}$, where $i$ becomes a Minkowski index ranging from 0 to 3. (We use the symbol $\hat =$ to represent a correspondence between Rosenfeld's terminology and our own). His $e_k$  raises Minkowski indices. Also, to avoid confusion when considering specific components, we use a capital letter to represent contravariant coordinate objects. So $\,e_k h_{k,\mu} \hat = e^K_\mu $, and $E^\nu{}_K \hat = \,e_k h_{k,\mu}$ is the reciprical of  $k h_{k,\mu}$ and $ h^\nu{}_k e_k h_{k, \mu} \hat = \delta^\nu_\mu$ is the statement that 
$E^\nu_K e^K_\mu = \delta^\nu_\mu$.
  
The spin connections are defined as
\beq
\omega_{\mu I J} = E^\alpha_I  \nabla_\mu e_{J \alpha}, \label{omega}
\eeq
where
$$
 \nabla_\mu e_{J \alpha} := \partial_\mu e_{J \alpha} - \Gamma_{\alpha \mu}^\beta e_{J \beta}.
$$
Expanding the Christoffel symbols in terms of the tetrads we find
\beq
\omega_\mu{}^{I J} = E^{\alpha I} e^J_{[\alpha, \mu]} - E^{\alpha J} e^I_{[\alpha, \mu]} + E^{\alpha I} E^{\beta J} e_{\mu L} e^L_{[\alpha, \beta]}. \label{Ricci}
\eeq 
 The curvature in terms of the spin connection is \footnote{Rosenfeld never explicitly referred to the spin coefficients.}
\beq
{}^{4}\!R_{\mu \nu}^{I J} =  \left( \partial_\mu \omega_\nu^{I J} - \partial_\nu \omega_\mu^{I J} +\omega_\mu{}^I{}_L \omega_\nu{}^{L J} -\omega_\nu{}^I{}_L \omega_\mu{}^{L J} \right). 
\eeq
Then the scalar curvature density is
\beq
{}^{4}\!{\cal R}= (-g)^\frac{1}{2} {}^{4}\!R 
= 2 (-g)^\frac{1}{2}E^\mu_I E^\nu_J \partial_\mu \omega_\nu^{I J}  +   (-g)^\frac{1}{2} E^\mu_I E^\nu_J \left(\omega_\mu{}^I{}_L \omega_\nu{}^{L J} -\omega_\nu{}^I{}_L \omega_\mu{}^{L J} \right).
\eeq
Rosenfeld took the gravitational Lagrangian density to be
\beq
{\cal L}_g =\frac{1}{2 \kappa}{\cal G} = \frac{1}{2 \kappa}(-g)^\frac{1}{2}E^\mu_I E^\nu_J  \left( \omega_\mu{}^I{}_L \omega_\nu{}^{L J} -\omega_\nu{}^I{}_L \omega_\mu{}^{L J} \right) \label{G},
\eeq
where $\kappa := 8 \pi G/c^2$. This is his expression (R104).
Thus Rosenfeld's gravitational Lagrangian is manifestly a scalar density under arbitrary coordinate transformations. It is also expressible as the sum of two manifest scalar densities,
\beq
{\cal L}_g = - \frac{1}{2 \kappa}{}^{4}\!{\cal R} + \frac{1}{ \kappa}\nabla_\mu \left[  (-g)^\frac{1}{2}E^\mu_I E^\nu_J  \omega_\nu^{I J}\right] =- \frac{1}{2 \kappa}{}^{4}\!{\cal R} +\frac{1}{ \kappa} \left[  E^{\mu I } \left((-g)^\frac{1}{2}E^\nu_I  \right)_{,\nu} \right]_{,\mu} . \label{gravdiv}
\eeq
This is the content of (R105). 
This quadratic Lagrangian is the analogue in terms of tetrads of the $\Gamma \Gamma$ Lagrangian employed originally by Einstein \cite{Einstein:1918ae} - with the significant difference that Einstein's $\Gamma \Gamma$ Lagrangian is not a scalar density. This introduced an extra complication in the Hamiltonian analysis of Bergmann and collaborators in the 1950's that was not present in Rosenfeld's model. On the other hand, Rosenfeld's Lagrangian is not invariant under local Lorentz transformations. Rosenfeld addressed this issue in his Case 2. 

\subsubsection*{The electromagnetic action}

We have the conventional electromagnetic action ${\cal E}$ in terms of the vector potential $A_\mu$ and field tensor $F_{\mu \nu} = A_{\nu, \mu} - A_{\mu, \nu}$,
\beq
{\cal E} = -\frac{1}{4} (-g)^{1/2} F_{\mu \nu} F^{\mu \nu}.
\eeq

\subsubsection*{The matter action}

Rosenfeld's matter Lagrangian is
\bea
 {\cal W} &=& \left( -g \right)^{1/2}\left[  \frac{1}{2}  i \bar \psi \gamma^\mu \left(\overrightarrow \partial_\mu +\Omega_\mu \right) \psi -\frac{1}{2} i\bar \psi  \left(\overleftarrow \partial_\mu -\Omega_\mu \right)\gamma^\mu  \psi   - m \bar \psi \psi  \right],
\eea
where $\gamma^\mu := E^\mu_I \Gamma^I$ and we denote the constant Dirac gamma matrices as $\Gamma^I$.
Also $\overline{\psi} := \psi^\dagger \Gamma^0$ and
\beq
\Omega_\mu  := \frac{1}{4} \Gamma^I \Gamma^J \omega_{\mu I J},
\eeq
is the spinor connection consistent with the Christoffel connection. It was first constructed independently by H. Weyl \cite{Weyl:1929ab} and V. Fock \cite{Fock:1929aa}. Both authors were attempting a geometric unification of Dirac's electron theory with gravity. \footnote{See the article by E. Scholz \cite{Scholz:2005aa}  for a discussion of the historical importance of this work both in the unification program and in the development of gauge theories in general. For the relevance to gauge theory see also the article by N. Straumann \cite{Straumann:2006ab}.}

We will use the properties
 \beq
 \Gamma_M \Gamma_N = -\Gamma_N \Gamma_M + 2 \eta_{MN},
 \eeq
 \beq
 \Gamma^\dagger_M = \Gamma^0\Gamma_M\Gamma^0,
 \eeq
 and
 \beq
 \left(\Gamma_M \Gamma_N\right)^\dagger =\Gamma^0 \Gamma_N \Gamma_M\Gamma^0,
 \eeq
 and hence
 \beq
 \Gamma^0 \Omega^\dagger_\mu \Gamma^0 = - \Omega_\mu.
 \eeq

\subsection{The momentae and the identities}

\subsubsection*{Case 1 - General covariance}

Rosenfeld's Case 1 assumes that the Lagrangian transforms as a scalar density under arbitrary coordinate transformations. This property is satisfied separately by ${\cal L}_g$, ${\cal E}$, and ${\cal W}$ under the transformations
\beq
x'^\mu = x^\mu + \xi^\mu(x).
\eeq
In particular, with ${\cal L} = {\cal L}_g + {\cal E}+ {\cal W} =:  {\cal L}_g +  {\cal L}_m$,
\beq
\delta {\cal L} + {\cal L} \xi^\mu_{,\mu} \equiv 0, \label{ident2}
\eeq
under the variations
\beq
\delta e_{\mu I} = - e_{\nu I} \xi^\nu_{, \mu} , \label{deltaegen}
\eeq
and
\beq
\delta A_\mu = - A_\nu \xi^\nu_{, \mu}. \label{deltaAgen}
\eeq
The identically vanishing coefficient of $\dot \xi^\mu$ in the identity (\ref{ident2}) then yields the four primary constraints
\beq
{\cal F}^I = p^{0 I} = 0, \label{constraint1}
\eeq
where
\beq
p^{\mu I} := \frac{\partial {\cal L}}{\partial \dot e_{\mu I}} 
\eeq
are the momenta conjugate to $e_{\mu I}$. Our (\ref{constraint1}) corresponds to (R117).

\subsubsection*{Case 1 - $U(1)$ Gauge invariance}

Rosenfeld's Case 1 also includes covariance under $U(1)$ transformations, $\delta x^\mu = 0$,
\beq
\delta A_\mu = \xi_{, \mu}, \label{deltaAu1}
\eeq
and
\beq
\delta \psi = i\frac{e}{\hbar c} \xi \psi, \label{deltapsiu1}
\eeq
under which $\delta {\cal L} = 0$. The coefficient of $\dot \xi$ in the identity (\ref{ident2}) then yields the additional primary constraint
\beq
{\cal F} = p^0 := \frac{\partial {\cal L}}{\partial \dot A_{0}} = 0, \label{constraint2}
\eeq
which is (R108). In the second line of (R109) Rosenfeld in principle displays two additional primary constraints, namely
\beq
p_\psi  = \left( -g \right)^{1/2} \frac{1}{2}  i \bar \psi \gamma^0, \label{ppsi}
\eeq
and
\beq
p_{\bar \psi} =- \left( -g \right)^{1/2} \frac{1}{2}  i   \gamma^0 \psi, \label{ppsid}
\eeq
which he, however, did not include among his ``eigentliche Identit\"aten" (proper identities). We will return to this neglect at the end of this section.

\subsubsection*{Case 2 - Local Lorentz invariance}

In Rosenfeld's Case 2 the symmetry variation of the Lagrangian picks up a total derivative, as is the case for local Lorentz transformations in Rosenfeld's Lagrangian. The matter and electromagnetic Lagrangians are invariant. But under 
the transformations with descriptors $\xi^{IJ} = -\xi^{JI}$,
\beq
 \delta e_{\mu M} =\eta_{MI} \xi^{IJ} e_{\mu J}, \label{deltaeLor}
\eeq
and
\beq
\delta \psi = \frac{1}{4} \xi^{IJ} \Gamma_I \Gamma_J  \psi, \label{deltapsiLor}
\eeq
we find that
\beq
 \delta {\cal L} =-  \frac{1}{\kappa}  \left[ \left(E^{\mu}_I E^\nu_J (-g)^\frac{1}{2}\right)_{,\nu}\xi^{IJ}  \right]_{,\mu}. \label{ident3}                
\eeq
Referring to (\ref{ident1aa}) we read off from this expression that
\beq
{\cal I}^\mu{}_{[IJ]} =   \frac{1}{\kappa} \left(E^{\mu}_{[I} E^\nu_{J]} (-g)^\frac{1}{2}\right)_{,\nu},  
\eeq
and according to (\ref{pcr}) the corresponding primary constraints are
\beq
{\cal F}'_{[IJ]} := p^{\mu}_{ [I} e_{J]\mu} + p_\psi \frac{1}{4} \Gamma_{[I }\Gamma_{J]} \psi + \frac{1}{4} \bar \psi\Gamma_{[I }\Gamma_{J]}  p_{\bar \psi} + \frac{1}{\kappa} \left(  \left((-g)^{1/2} E^0_{[I} E^a_{J]}  \right)_{, a} \right) = 0. \label{constraint3}
\eeq
This is not exactly (R124) because Rosenfeld took the constraints (\ref{ppsi}) and (\ref{ppsid}) as identities. If these are inserted into (\ref{constraint3}) we obtain (R124).

 As we pointed out earlier, unless Rosenfeld indicates otherwise,  he conceived all of his variables as quantum mechanical operators. And although he does not say so explicitly, we are to understand that the spinorial variables are to satisfy the anti-commutation relation 
\beq
\left[\psi_\rho, \psi'_\rho\right]_+ := \psi_\rho \psi'_\rho + \psi'_\rho \psi_\rho = \delta_{\rho \sigma} \delta^3 (\vec x, \vec x'), \label{algebra}
\eeq
as in (57a) in the foundational quantum field article by Heisenberg and Pauli \cite{Heisenberg:1929aa} that served as Rosenfeld's inspiration for this paper. This interpretation is consistent with Rosenfeld's footnote following his equation (R107) in which he mentions that it is not necessary in this article to discuss the modifications in the general scheme that the use of spinorial variables entails. We make this point since from the point of view of later developments in constrained Hamiltonian dynamics, the constraints  (\ref{ppsi}) and (\ref{ppsid}) are ``second class''; they do not have vanishing Poisson brackets with all of the constraints. Following the procedure later introduced independently by Bergmann and by Dirac, new Poisson brackets need to be constructed that respect the constraints. However, it turns out that in this case the new Poisson brackets correspond precisely to the quantum anti-commutation relations employed by Heisenberg and Pauli - and it is legitimate to use these relations in computing the action of all of the operators that are exhibited by Rosenfeld in this paper.

\subsection{Symmetry generators}

Next we construct the symmetry variations according to paragraphs \S 13 and \S 14. Substituting the variations (\ref{deltaegen}), (\ref{deltaAgen}), (\ref{deltaAu1}), (\ref{deltapsiu1}),(\ref{deltaeLor}), and (\ref{deltapsiLor}) into the Rosenfeld-Noether generator density in (\ref{M0}) we obtain according to (R59)
\bea
{\cal M}^0 &=& -{\cal F}^I e_{0I} \dot \xi^0  -{\cal F}^I e_{aI} \dot \xi^a - {\cal F} A_0 \dot \xi^0-p^{a I}  e_{\nu I} \xi^\nu_{, a} - p^a A_{\nu} \xi^\nu_{,a}- {\cal H} A_0 \xi^0 - {\cal G}_a \xi^a \nonumber \\
&-&{\cal F} \dot \xi +  p^a \xi_{,a} + i \frac{e}{\hbar c} p_\psi \psi \xi-  i \frac{e}{\hbar c}p_{\psi^\dagger} \psi^\dagger \xi + {\cal F}'_{[IJ]} \xi^{IJ}.
\eea
We are assured, following the proof in section 3, that the charge $M[\xi] :=\int d^3\!x {\cal M}^0$ generates the correct symmetry variations of all of the phase space variables.

Also, according to the arguments in section 3, we obtain the following secondary constraints,
\beq
{\cal N}^0_0 := \left(p^{a I}  e_{0 I}\right)_{,a}  +\left( p^a A_{0} \right)_{,a}- {\cal H} = 0, \label{hconstraint}
\eeq
\beq
{\cal N}^0_b:= \left(p^{a I}  e_{b I}\right)_{,a}  - {\cal G}_b = 0, \label{gconstraint}
\eeq
and
\beq
{\cal N}^0:=-p^a_{,a} + i \frac{e}{\hbar c} p_\psi \psi -  i \frac{e}{\hbar c}p_{\psi^\dagger} \psi^\dagger = 0.
\eeq

The secondary constraints (\ref{hconstraint}) and (\ref{gconstraint}) appear in Rosenfeld's article for the first time as the phase space expression of the four Einstein equations that do not involve accelerations. Following their appearance in (R119)  he noted that Klein had obtained them in another context in 1918 \cite{Klein:1918aa}. In fact, Klein obtained them in the context of the $\Gamma \Gamma$ gravitational Lagrangian. This lends support to our claim that had he wished, Rosenfeld could easily have extended his method to the another case that was pursued by Schiller in his Ph. D. thesis.

Let us write the Rosenfeld-Noether generator $M[\xi]$ as the sum
$$
 M[\xi^\mu, \xi, \xi^{IJ}] =D[\xi^\mu]+U[\xi]+L[\xi^{IJ}],
 $$
thereby distinguishing the generators for general coordinate transformations, $U(1)$ gauge transformations and local Lorentz transformations. We obtain, by introducing the secondary constraints (\ref{hconstraint}, \ref{gconstraint})
\bea
D[\xi^\mu] &= &\int d^3\!x \left( -{\cal F}^I e_{0I} \dot \xi^0  -{\cal F}^I e_{aI} \dot \xi^a - {\cal F} A_0 \dot \xi^0-p^{a I}  e_{\nu I} \xi^\nu_{, a} - p^a A_{\nu} \xi^\nu_{,a}- {\cal H} A_0 \xi^0 - {\cal G}_a \xi^a\right) \nonumber \\
&= & \int d^3\!x \left(-{\cal F}^I e_{0I} \dot \xi^0  -{\cal F}^I e_{aI} \dot \xi^a - {\cal F} A_0 \dot \xi^0+{\cal N}^0_0 \xi^0 + {\cal N}^0_a \xi^a \right) +b.t. \nonumber  \\
&= & \int d^3\!x \left((-{\cal F}^I e_{0I} - {\cal F} A_0)\dot \xi^0  -{\cal F}^I e_{aI} \dot \xi^a +{\cal N}^0_0 \xi^0 + {\cal N}^0_a \xi^a \right) + b.t., \label{Dgen}
\eea
where {\it b.t.} denotes a boundary term.

Next, the generator for infinitesimal $U(1)$ transformations is
\bea
U[\xi] &= &\int d^3\!x \left(-{\cal F} \dot \xi +  p^a \xi_{,a} + i \frac{e}{\hbar c} p_\psi \psi \xi-  i \frac{e}{\hbar c}p_{\psi^\dagger} \psi^\dagger \xi\right) \nonumber \\
&= & \int d^3\!x \left(-{\cal F} \dot \xi  + {\cal N}^0 \xi \right) +b.t. \label{Ggen}
\eea
It may be  rewritten further by using the relations (\ref{ppsi}) and (\ref{ppsid}) and then employing (as did Rosenfeld) the algebra (\ref{algebra}), 
\beq
{\cal N}^0 = -p^a_{,a} + \frac{e}{\hbar c} \left( -g \right)^{1/2}   \psi^\dagger \psi = 0.
\eeq
Thus we have here the first published derivation of Gauss' law in a constrained Hamiltonian formalism.

Finally, the generator of infinitesimal local Lorentz transformations is
\beq
L[\xi^{IJ}] = \int d^3\!x {\cal F}'_{[IJ]} \xi^{IJ}. \label{Lgen}
\eeq

One knows that for a variety of examples (see e.g. \cite{Sundermeyer:2014aa}), that for local symmetries characterized by descriptors $\xi^{A}$, the gauge generator has the form
$$
G[\xi^{A}] = \int d^3\!x \,(\phi_A \xi^{A} + \psi_A\dot{\xi}^{A})
$$
where the $\{\phi_A, \psi_A \}$ are all first-class constraints. The expressions (\ref{Dgen}, \ref{Ggen}, \ref{Lgen}) have this generic structure, although the first class property need not hold for the coefficients of the $\xi^A$ and $\dot \xi^A$ in the Rosenfeld-Noether generators above. Indeed, we expect that one needs to work with modified transformations in order to respect the Legendre projectability of the transformations \cite{Pons:2000aa}. These questions will be addressed in a future publication.

\subsection{Finite canonical transformations and the symmetry group algebra}

A subset of Rosenfeld's infinitesimal symmetry transformations can be readily extended to finite transformations. So, for example, a finite Lorentz rotation with finite descriptor $\xi^{IJ}$ is generated by the exponentiated generator $M_L$, defined as a sum of nested Poisson brackets,
\beq
\exp \left(M_L\right) := 1 + \left\{ \ldots , M_L\right\} + \frac{1}{2} \left\{ \left\{ \ldots , M_L\right\} , M_L \right\} + \ldots
\eeq

One can also realize canonical active finite 3-D diffeomorphism transformations. Suppose, for example, that we wish to actively transform a scalar phase space function $\phi(x)$ under the diffeomorphism $x'^a = x^a + \xi^a(\vec x)$. Its finite actively transformed change is then
\beq
\phi' - \phi = \phi - \phi_{,a} \xi^a + \frac{1}{2}\left( \phi_{,a} \xi^a \right)_{,b} \xi^b + \ldots
\eeq
This is indeed generated by $N\left[ \vec \xi \right]:=\int d^3\!x {\cal N}^0_a \xi^a$,
\beq
\phi' =\exp\left( N\left[ \vec \xi \right] \right)\phi = \phi + \left\{  \phi, N\left[ \vec \xi \right]\right\} + \frac{1}{2}\left\{ \left\{  \phi, N\left[ \vec \xi \right]\right\}, N\left[ \vec \xi \right] \right\} + \ldots
\eeq

However, we cannot realize arbitrary 4-D diffeomorphisms, $x'^\mu = x^\mu + \xi^\mu( x)$ with $x^0 \ne 0$, in this way. Rosenfeld eliminated precisely these transformations in his derivation of the group algebra in his equation (R53).  \footnote{The closure of this phase space algebra will later be the defining property of first class constraints \cite{Dirac:1950aa}. Curiously, Dirac introduced this notion without ever mentioning the work of Rosenfeld with which he had been familiar since 1932. See \cite{Salisbury:2009ab} for a discussion of relevant correspondence in 1932 between Rosenfeld and Dirac.} The reason is that time derivatives up to infinite order of the descriptors $\xi^\mu$ appear in the actively transformed phase space functions, and they do not appear in $M\left[ \xi \right]$. One encounters a related difficulty in considering the commutator of infinitesmal transformations. If we undertake the infinitesimal transformation $x_1^\mu = x^\mu + \xi_1^\mu( x)$ followed by $x_2^\mu = x^\mu + \xi_2^\mu( x)$ and then subtract the them in reverse order, the descriptor of the overall coordinate transformation is
\beq
\xi^\mu_3 = \xi^\mu_{1, \nu} \xi_2^\nu - \xi^\mu_{2, \nu} \xi_1^\nu.
\eeq
Thus higher order time derivatives appear with each commutation. The appearance of these time derivatives was an early concern of Bergmann - although not explicitly stated by him or his collaborators in the period prior to Dirac's gravitational Hamiltonian breakthrough in 1958 \cite{Dirac:1958aa}. He did however refer to this challenge in a later recollection\footnote{``During the early Fifties those of us interested in a Hamiltonian
formulation of general relativity were frustrated by a
recognition that no possible canonical transformations of the field variables could mirror four-dimensional coordinate transformations
and their commutators, not even at the infinitesimal level. That
is because (infinitesimal or finite) canonical transformations
deal with the dynamical variables on a three-dimensional hypersurface,
a Cauchy surface, and the commutator of two such infinitesimal
transformations must be an infinitesimal transformation
of the same kind. However, the commutator of two infinitesimal
diffeomorphisms involves not only the data on a three-dimensional
hypersurface but their "time"-derivatives as well. And if these
data be added to those drawn on initially, then, in order to obtain
first-order "time" derivatives of the commutator, one requires
second-order "time" derivatives of the two commutating diffeomorphisms,
and so forth. The Lie algebra simply will not close."\cite{Bergmann:1979aa}, pp. 174-175 }

We can surmise from Rosenfeld's discussion in his section \S 6 that he must have recognized this obstacle since as we noted above he confined his discussion of the group algebra to the spatial diffeomorphisms and the internal symmetry transformations. He concluded that the vanishing generators must satisfy a closed Lie algebra. In the language that was introduced later by Dirac  \cite{Dirac:1950aa} (without ever mentioning the work of Rosenfeld with which he had been familiar since 1932)\footnote{See \cite{Salisbury:2009ab} for a discussion of relevant correspondence in 1932 between Rosenfeld and Dirac.}, the constraints must be first class.

The impossibility of generating finite canonical transformations corresponding to coordinate transformations for which $\delta x^0 \neq 0$ was a feature not only of Rosenfeld's Noether charge, but also the generators that were written down first by Anderson and Bergmann\cite{Anderson:1951aa} in 1951, and later by Bergmann and Schiller \cite{Bergmann:1953aa} in 1953.  Although Dirac never concerned himself with the question whether the full diffeomorphism group could be realized as a canonical transformation group, he is the one who unintentionally invented the framework in which this goal could be achieved. The key was the decomposition of infinitesimal coordinate transformations which were either tangent to a given foliation of spacetime into fixed time slices, or perpendicular to the foliation. Bergmann and Komar \cite{Bergmann:1972aa} subsequently gave a group-theoretical interpretation of this decomposition, pointing out that the relevant group was a phase space transformation group that possessed a compulsory dependence on the spacetime metric. In 1983 we \cite{Salisbury:1983aa} provided an explicit proof that this dependence was required in order to obtain a Lie algebra that did not involve higher time derivatives of the descriptors. More recently, Pons, Salisbury and Shepley \cite{Pons:1997aa} showed that this demand on the structure of the group algebra is equivalent to the demand that the permissible variations of configuration-velocity variables be projectable under the Legendre transformation from configuration-velocity space to phase space.

\subsection{Expanding upon Rosenfeld's application: Construction of the Hamiltonian}

Because of the existence of primary constraints it is not possible to solve uniquely for the momentae in terms of the velocities. As we noted earlier, Rosenfeld pioneered a method for obtaining general solutions that involved as many arbitrary functions as there were primary constraints, where the constraints arising in both Cases 1 and 2 must be taken into account. Rosenfeld then employed these general solutions in the construction of the Hamiltonian.

Rosenfeld did not display the explicit expression for the Hamiltonian for his general relativistic model. We do not know why. It is, however, straightforward to apply his method to construct it. We undertake the construction here.

We begin with the momentum conjugate to the tetrads $e_{\mu I}$, (R118), 
\bea
  &&p^{\mu I}=  \frac{\partial {\cal L}_g }{\partial \dot e_{\mu I}} + \frac{\partial {\cal L}_m }{\partial \dot e_{\mu I}} \nonumber \\
  &=& \frac{1}{\kappa}(-g)^{1/2} \left( - 2 g^{\beta [ 0} E^{\mu ]I} E^\alpha_M +  g^{\alpha [ \mu} E^{0 ]}_M E^{\beta I} + \frac{1}{2}g^{\alpha [ \mu} g^{0 ] \beta} \delta^I_M \right) \left( e^M_{\alpha , \beta} - e^M_{\beta, \alpha} \right) \nonumber \\
  &+&\frac{1}{8}i \bar \psi\left( -g \right)^{1/2}\left[    \left(\gamma^\nu \Gamma_K \Gamma_J +\Gamma_K \Gamma_J\gamma^\nu \right) \left(2 \eta^{I[J} E^{K] [\mu} \delta^{0]}_\nu +E^{\mu [K} E^{J]0} e^I_{\nu}\right)  \right]\psi \nonumber \\ \label{pmuI}
\eea
In seeking appropriate linear combinations of the velocities that can be solved in terms of the momentae, it seems natural to define the symmetric and antisymmetric combinations
\beq
s_{(\mu \nu)} :=   e^I_{(\mu} \dot e_{\nu) I} =   e_{(\mu} \cdot \dot e_{\nu)} = \frac{1}{2}\dot g_{\mu \nu},
\eeq
and
\beq
a_{[\mu \nu]}: =  e_{[\mu} \cdot \dot e_{\nu] }.
\eeq
It does turn out that the $a_{[\mu \nu]}$ combinations do not appear in (\ref{pmuI}), and neither do the $s_{(0 \mu)}$.  Defining $S^{(\alpha \beta)} := E^{I(\alpha} p^{\beta)}_I$ and $A^{[\alpha \beta]} := E^{I[\alpha} p^{\beta]}_I$ we obtain the relations
\beq
 S^{(\mu 0)} =  N^{(\mu 0)}, \label{Smu}
\eeq
and
\beq
 A^{[\mu \nu]} = N^{[\mu \nu]}, \label{Amunu}
\eeq
plus the six linear equations for the $s_{(a b)}$,
\beq
E^{(a } \cdot p^{b)} = M^{(ab) (cd)} s_{(cd)} + N^{(ab)}, \label{s}
\eeq
where
\bea
N^{[\mu \nu]} &=&\frac{1}{2} (-g)^{1/2}\left( g^{\alpha [ \mu} g^{\nu] b}E^{0 }_M + g^{b [\mu} g^{\nu] 0} E^\alpha_M   - g^{\alpha [\mu} g^{\nu] 0} E^b_M   \right) e^M_{\alpha , b} \nonumber \\
&+& \frac{1}{8}i \bar \psi\left( -g \right)^{1/2}   \left(\gamma^\rho \Gamma_{[K} \Gamma_{J]} +\Gamma_{[K} \Gamma_{J]}\gamma^\rho \right) E^{J \nu } E^{ \mu K} \delta^{0}_\rho  \psi,
\eea
\bea
N^{(\mu 0)}&=&\frac{1}{2}(-g)^{1/2}\left( -  g^{c  0} g^{\mu 0} E^\alpha_M   + g^{\alpha  (\mu} E^{0)}_Mg^{0 c}  \right) e^M_{\alpha , c}  \nonumber \\
&+&\left(   g^{c (\mu} g^{0) 0} E^\alpha_M -  g^{\alpha 0} E^{(\mu}_M g^{0) c}  \right) e^M_{\alpha , c}  \nonumber \\
&+& \left(    g^{\alpha  0} g^{\mu 0} E^c_M   - g^{\alpha (\mu} g^{0) 0} E^c_M  \right) e^M_{\alpha , c} \nonumber \\
&+& \frac{1}{4}i \bar \psi\left( -g \right)^{1/2}    \left(\gamma^\rho \Gamma_{[K} \Gamma_{J]} +\Gamma_{[K} \Gamma_{J]}\gamma^\rho \right) E^{K (0}  \delta^{\nu)}_\rho E^{J0} \psi,
\eea
\bea
N^{(ab)} &=& \nonumber \\
&+& \frac{1}{4}i \bar \psi\left( -g \right)^{1/2}    \left(\gamma^\rho \Gamma_{[K} \Gamma_{J]} +\Gamma_{[K} \Gamma_{J]}\gamma^\rho \right) E^{K (a}  \delta^{b)}_\rho E^{J0} \psi,
\eea
and
\beq
M^{(ab) (cd)} =\frac{1}{2}\sqrt{{}^3\!g} N^{-1} \left(- 2  e^{a b} e^{d c}  + e^{d  a}  e^{c b}  +    e^{a c} e^{d b}  \right).
\eeq

The relation (\ref{Amunu}) contracted with $e_{\mu I} e_{\nu J}$ is the primary constraint (\ref{constraint3}). Also the contraction of (\ref{Smu}) with $e_{\mu J}$ is the primary constraint (\ref{constraint2}).

We now solve (\ref{s}) for the $s_{(cd)}$, obtaining what Rosenfeld calls the special solutions. Employing Rosenfeld's notation, we label the special solutions with a superscript $0$.
Then using the inverse of $M^{(ab) (cd)} $ we find
\bea
s^0_{(cd)} &=& M_{ (cd) (ab)} \left(  E^{(a } \cdot p^{b)} - N^{(ab)} \right) \nonumber \\
&=&  \frac{N}{ 2\sqrt{{}^3\!g}}\left(-g_{ab} g_{cd} + g_{ac} g_{bd} +  g_{ad} g_{bc}  \right) \left(  E^{(a } \cdot p^{b)} - N^{(ab)} \right)  \label{ysol}
\eea
Following Rosenfeld (see (R32) and (\ref{32})), we also know that we can take
\beq
a^0_{[\mu \nu]} = 0,
\eeq
and
\beq
s^0_{(0 \mu)} = 0.
\eeq

Again following Rosenfeld the general solutions for the velocities are therefore
\beq
\dot e_{\mu I} = E^\nu_I \left( s^0_{(\nu \mu)} +a^0_{[\nu \mu]}+  \right) = \delta_\mu^b E^a_I s^0_{(a b)} +E^\nu_I\left( \lambda_\nu\delta^0_\mu + \lambda_{[\mu \nu]} \right), \label{general}
\eeq
where $\lambda_\nu$ and $\lambda_{[\mu \nu]}$ are arbitrary spacetime functions. As Rosenfeld noted, these arbitrary functions do not appear when these velocities are substituted into the Lagrangian. On the other hand, defining the arbitrary functions $\lambda_I :=E^\nu_I \lambda_\nu$ and $\lambda_{[IJ]} := E^\mu_I E^\nu_J \lambda_{[\mu \nu]}$ we find that
\beq
p^{\mu I}  E^\nu_I\left( \lambda_\nu\delta^0_\mu + \lambda_{[\mu \nu]} \right) = : \lambda_I {\cal F}^I + \lambda^{[IJ]} {\cal F}'_{[IJ]}.
\eeq

Thus substituting the general solution (\ref{general}) into $p^{\mu I} \dot e_{\mu I} - {\cal L}$ we obtain the purely gravitational contribution to the Hamiltonian density
\beq
{\cal H}_g = {\cal H}^c_g +  \lambda_I {\cal F}^I + \lambda^{[IJ]} {\cal F}'_{[IJ]} \label{hg}
\eeq
with
$$
{\cal H}^c_g = \frac{1}{2} S^{(ab)} M_{ (ab)(cd)} S^{(cd)} - N^{(ab)} M_{ (ab)(cd)} S^{(cd)} + \frac{1}{2} N^{(ab)} M_{ (ab)(cd)} N^{(cd)} -{\cal A}
$$
where $$
{\cal A} = \frac{1}{8 \kappa} \left(-g\right)^{1/2} \left( 4 E^{[\alpha}_M g^{a] [b}E^{\rho]}_N- 2 E^{[a}_N g^{\alpha][\rho} E^{b]}_M - \eta_{MN} g^{\rho [\alpha} g^{a]b} \right)e^M_{\alpha,a} e^N_{\rho,b}
$$
 is the velocity-independent term in ${\cal L}_g$.

The total Hamiltonian density is ${\cal H}_g + {\cal H}_{em}$, where the electromagnetic contribution ${\cal H}_{em}$ can also be found applying Rosenfeld's method. See \cite{Salisbury:2009ab}. It has the structure ${\cal H}_{em} = {\cal H}^c_{em} + \lambda {\cal F} + \lambda^{i} \chi_i$ where the $\chi_i$ are the spinorial constraints from (\ref{ppsi}, \ref{ppsid}). In the "usual" Dirac-Bergmann procedure one would require the stabilization of primary constraints, and/or find new constraints or fix the multipliers $\lambda$. Although the Hamiltonian (\ref{hg}) generates the correct field equations, some additional work needs to be done to be able to compare with later publications on canonical tetrad-spinor formulations, as for instance  \cite{Nelson:1977aa,Nelson:1978aa}.
The task would be to check that ${\cal H}^c_{g} + {\cal H}^c_{em}$ is a linear combination of secondary constraints such the total Hamiltonian vanishes weakly (up to a total divergence) - as expected for generally covariant theories.

\section{Conclusions}

L\'eon Rosenfeld's 1930 {\it Annalen der Physik} paper not only developed a comprehensive Hamiltonian theory to deal with local symmetries that arise in Lagrangian field theory, but he already disclosed connections between symmetries, constraints, and phase space symmetry generators. Indeed, to a surprising degree he established the foundational principles that would later be rediscovered and in some respects extended by the individuals who until recently have been recognized as the inventors of the methods of constrained Hamiltonian dynamics, Peter Bergmann and Paul Dirac. Not only did he provide the tools to deal with the only local gauge symmetries that were known at the time, namely local $U(1)$ and local Lorentz covariance, but perhaps more importantly he also established the technique for translating into a Hamiltonian description the general covariance under arbitrary spacetime coordinate transformations of Einstein's general theory of relativity. Some of this pioneering work either became known or was independently rediscovered over two decades later. But for unknown reasons Rosenfeld never claimed ownership, nor did he join later efforts to exploit his techniques in pursuing canonical approaches to quantum gravity.

It is remarkable that Rosenfeld's article remained unknown to the community. Even the most cited monographs on constrained dynamics \cite{Hanson:1976aa,Henneaux:1992aa,Sundermeyer:1982aa}
omit Rosenfeld's article\footnote{\,\,\,The present article may thus be seen as an atonement to Rosenfeld by one of the authors.}. Why did this happen? It seems likely that Pauli's lack of appreciation and/or understanding could have influenced Rosenfeld's decision not to promote his work. We get a sense of Pauli's attitude from a letter written by Pauli to Oskar Klein in 1955: ``I would like to bring to your attention the work by Rosenfeld in 1930. He was known here at the time as the Ôman who quantised the Vierbein (sounds like the title of a Grimms fairy tale doesnÕt it?). See part II of his work where the Vierbein
appears. Much importance was given at that time to the identities among the p's and q's (that is the canonically conjugate fields) that arise from the existence of the group of general coordinate transformations. I still remember that I was not happy with every aspect of his work since he had to introduce certain additional assumptions that no one was satisfied with. " \footnote{ÔÔGerne m\"ochte ich Dich in dieser Verbindung auf die lange Arbeit von Rosenfeld, Annalen der Physik (4), 5, 113, 1930 aufmerksam machen. Er hat sie seinerzeit bei mir in Z\"urich gemacht und hiess hier dementsprechend Ôder Mann, der das Vierbein quanteltÕ (klingt wie der Titel eines Grimmschen M\"archens, nicht?). Siehe dazu Teil II seiner Arbeit, wo das ÔVierbeinÕ daran kommt. Auf die Identit\"aten zwischen den ÔpÕ und ÔqÕ - d.h. kanonisch konjugierten Feldern \, die eben aus der Existenz der Gruppe der Allgemeinen Relativit\"atstheorie (Koordinaten Transformationen mit 4 willk\"urlichen Funktionen) entspringen, wurde damals besonderer Wert gelegt. Ich erinnere mich noch, dass Rosenfelds Arbeit nicht in jeder Hinsicht befriedigend war, da er gewisse zus\"atzliche Bedingungen einf\"uhren musste, die niemand richtig verstehen konnteÕÕ \cite{Meyenn:2001aa}, p. 64.} Indeed, as we have shown, it only became apparent in his Part 2 that the special cases that Rosenfeld identified in his Part 1 were chosen with the Einstein-Maxwell-Dirac theory in mind, and the article might have been more accessible had he simply addressed this model from the start rather than formally treating a wider class of theories. It is this lament by Pauli that leads us to suspect that Rosenfeld's general theory was indeed more general than the unidentified Pauli suggestion that Rosenfeld acknowledged in his introduction. 

Yet the paper was known, in particular already in 1932 by  Dirac, as has been documented elsewhere \cite{Salisbury:2009ab}, yet Dirac did not cite it in his papers on constrained Hamiltonian dynamics \cite{Dirac:1950aa,Dirac:1951aa}. Strangely, in another paper of 1951 concerned with electromagnetism in flat spacetime Dirac did refer to Rosenfeld in addition to his foundational papers in declaring that ``an old method of Rosenfeld (1930) is adequate in this case" in making the transition from a Lagrangian to the Hamiltonian.\footnote{\cite{Dirac:1951ab}, p. 293 } With regard to the Syracuse group, the paper was only discovered following the publications by Bergmann \cite{Bergmann:1949aa} and Bergmann-Brunings \cite{Bergmann:1949ab} of their initial foundational papers on constrained Hamiltonian dynamics.\footnote{J. Anderson related to D. S. in 2006 that it was he who had found the paper and brought it to the attention of Bergmann. In this same conversation R. Schiller indicated that the paper was the inspiration for his Ph. D. thesis, conducted under Bergmann's direction.} As we noted earlier, following this discovery Schiller made explicit use of the Rosenfeld paper in constructing the phase space generators of symmetry transformations that we have elected to call Rosenfeld-Noether generators. On the other hand, in the joint publication by Bergmann and Schiller \cite{Bergmann:1953aa} that focused on these charges Rosenfeld was not cited.

Rosenfeld's article begins with general discussion regarding the consequences of local symmetries existing in the in the cotangent bundle space of symmetries in configuration-velocity space. He not only (1) derives identities following from the invariance of a Lagrangian and uses them for obtaining phase-space constraints, but he also (2) proposes an expression for the generator of phase-space symmetry transformations, and (3) details a procedure to derive a Hamiltonian density from a singular Lagrangian in a manner more mathematically satisfying  than later ones by Dirac and by Bergmann and his Syracuse group.

The history-of-science story of the Klein-Noether identities is another story of early discovery and later rediscovery. Felix Klein in 1918 derived a chain of identities for general relativity in his attempt to arrive at conservation laws in general relativity \cite{Klein:1918aa}. Similar chains of identities exist for arbitrary local symmetries; they shall not be derived here (for details see Sect. 3.3.3 in \cite{Sundermeyer:2014aa}) These identities have as a consequence what is known as Noether identities, namely identically fulfilled relations involving the Euler derivatives and derivatives thereof. Another consequence is the vanishing of the Hessian determinant, which is a characteristic of a singular Lagrangian with ensuing phase-space constraints. 

The full set of Klein-Noether identities was investigated also by J. Goldberg \cite{Goldberg:1953aa}, exhibited by R. Utiyama \cite{Utiyama:1959aa}, mentioned by A. Trautman \cite{Trautman:1967ab} - all of them not citing F. Klein. (It seems that the first reference to Klein is in \cite{Barbashov:1983aa}.) The identities were called extended Noether identities in  \cite{Lusanna:1990aa,Lusanna:1991aa} , cascade equations in \cite{Julia:1998aa,Julia:2000aa}, Noether's third theorem in \cite{Brading:2003aa,Brading:2002aa}, and Klein identities in \cite{Petrov:2013aa}.

Another result concerning the Klein-Noether identities - already visible in the Rosenfeld article, and still widely unknown today - is the fact that these are entirely equivalent to the chain of primary, secondary, ... constraints in the Hamiltonian treatment  \cite{Lusanna:1990aa,Lusanna:1991aa}.

And still another history-of-science story lays dormant under repeated efforts to find generators of phase-space symmetry transformations. After Rosenfeld, the investigations into the manner in which the constraints of a theory with local symmetries relate to the generators of these symmetries in phase space restarted with the work of Anderson and Bergmann \cite{Anderson:1951aa}, Dirac \cite{Dirac:1964aa}, and Mukunda  \cite{Mukunda:1976aa,Mukunda:1980aa}. It soon became clear that the phase space symmetry generator is a specific linear combination of the first-class constraints. In 1982, Castellani devised an algorithm to determine a symmetry generator \cite{Castellani:1982aa}. This was completed by Pons/Salisbury/Shepley \cite{Pons:1997aa} by taking  Legendre projectability into account and thereby extending the formalism to incorporate finite symmetry transformations.

It seems to have gone unnoticed that L. Rosenfeld already in 1930 showed that the vanishing charge associated with the conserved Noether symmetry current is the sought-after phase-space symmetry generator, called the Rosenfeld-Noether generator in this article. The figure who came the closest to affirming this fact was L. Lusanna who indeed contemplated a wider scope of symmetry transformations including several specific pathological cases \cite{Lusanna:1990aa,Lusanna:1991aa}. The proof by Rosenfeld, repeated in Section 4, is not easy to digest at first reading, but it is valid for infinitesimal transformation. One consequence that all examples suggest is that one can read off the first-class constraints of the theory in question from the Rosenfeld-Noether generator. Recall that in the "usual" handling of constrained systems, sometimes referred to as the Dirac-Bergmann algorithm, one needs to establish a Hamiltonian first in order to find all constraints beyond the primary constraints. Only then can first and second-class objects can be defined. 

In a forthcoming article we will show how Rosenfeld's approach can be generalized so that Legendre projectability is respected. 
One significant result of this analysis is that whenever local symmetries beyond general covariance are present, such appropriately chosen symmetries must be added to the general coordinate transformations to achieve canonically realizable transformations.\footnote{See e.g. \cite{Pons:2000aa} and \cite{Pons:2000ab} for a discussion of these additions in the context of Einstein-Yang-Mills and the Ashtekar formulation of general relativity.}

With his attempt to quantize the Einstein-Maxwell-Dirac theory Rosenfeld made an ambitious effort that was ``well before its time".  
Keep in mind that prior to Rosenfeld's article no results on the Hamiltonian formulation of pure Einstein gravity were known, that Weyl's ideas of electromagnetic gauge invariance were not generally accepted, and that spinorial entities were still treated {\it ad hoc}. He can be forgiven for not having reached today's level of understanding. He did not derive explicitly all of the first-class constraints from the Klein-Noether identities although he did appreciate their importance as group generators. Nor did he display the full Hamiltonian for his model even though as we have seen he was certainly in position to do so in a straightforward application of his method. Thus he could have derived a tetrad formulation for general relativity with gauge fields nearly five decades before it appeared on the quantum gravitational research agenda.

As a matter of fact the canonical formulation of general relativity in terms of tetrads and spin connections became a hot topic only in the 1970's - even though Bryce DeWitt and Cecile DeWitt-Morette had addressed this issue already in 1952 \cite{DeWitt:1952ab}. The preponderance of articles on canonical general relativity around 1950 were formulated in terms of the metric and the Levi-Civita connection. Rosenfeld obviously was aware that this was possible in the case of vacuum general relativity. In item (3) of his $\S$15 he writes ``The pure (vacuum) gravitational field could be described by the $g_{\mu \nu}$ instead of the $h_{i, \nu}$. Then we would be dealing with another variation of the `second case' ''. Indeed, he notes that as a consequence of the general covariance four primary constraints would arise (that first appeared explicitly in Bergmann and Anderson). It would be of interest to apply a modified version of Rosenfeld's program to both the Dirac \cite{Dirac:1958aa} and to the ADM \cite{Arnowitt:1962aa} Lagrangians. These differ by divergence terms.\footnote{See e.g. \cite{Cianfrani:2012aa}} The divergence terms do not however transform as scalar densities under general coordinate transformations, so their treatment would require a simple modification of Rosenfeld's Case two.

Of course, Ashtekar's invention of new gravitational variables initiated an interest in tetrad variables that form the basis of today's active research in loop quantum gravity. And we can thank Rosenfeld for not only setting down the first stones of the foundations for this canonical loop approach to quantum gravity. Remarkably, in addition he pioneered the development of the gauge theoretical phase space framework that undergirds all current efforts at unifying the fundamental physical interactions.

\appendix
\section{Constrained Dynamics}

\subsection{Singular Lagrangians}

Assume a classical theory with a finite number of degrees of freedom $q^k\,\,(k=1,...,N)$ defined by its Lagrange function $L(q,\dot{q})$ with the equations of motion
\begin{equation}
[L]_k\,\,:=\,\frac{\partial L}{\partial q^k}- \,\frac{d}{dt}\,\,\frac{\partial L}{\partial \dot{q}^k}
=\,\Big(\frac{\partial L}{\partial q^k}-\,\frac{\partial^2 L}{\partial \dot{q}^k \partial q^j} \dot{q}^j\Big)
-\,\frac{\partial^2 L}{\partial \dot{q}^k
\partial \dot{q}^j} \ddot{q}^j
=: V_k-W_{kj}\ddot{q}j =0. \label{hessEOM} 
\end{equation} 
For simplicity, it is assumed that the Lagrange function does not depend on time explicitly; all the following results can readily be extended. A crucial role is played by the matrix (sometimes called the "Hessian")
\beq 
W_{kj}\,:=\,\frac{\partial^2 L}{\partial \dot{q}^k \partial \dot{q}^j}.  \label{hessian} 
\eeq
If $\,\,\det W = 0,$ not only the Lagrangian but the system itself is termed 'singular', and 'regular' otherwise.

From the definition of momenta by 
\beq 
p_k(q,\dot{q})= \frac{\partial L}{\partial \dot{q}^k},
\label{momentum} 
\eeq 
one immediately observes that only in the regular case can the $\,p_k(q,\dot{q})\,$ be solved for all the velocities in the form $\,\dot{q}^j(q,p)\,$ - at least locally.

In the singular case, $\,\det W = 0\,$ implies that the $\,N\times N$ matrix $\,W$ has a rank $R\,$ smaller than $N\,$ - or that there are $\,P=N$-$R\,$ null eigenvectors $\,\xi^k_{\rho}\,$:
\begin{equation}
\label{null-ev}
\xi^k_{\rho}\,W_{kj}\equiv 0\hspace{1cm}\textrm{for}\hspace{1cm} \rho=1,...,P \,(= N-R).
\end{equation}
This rank is independent of which generalized coordinates are chosen for the Lagrange function. The null eigenvectors serve to identify those of the equations of motion which are not of second order. By contracting these with $\,\ddot{q}^j\,$ one gets the $\,P\,$ on-shell equations
$$
\chi_\rho=\xi^k_{\rho}\,V_{k}(q,\dot{q}) = 0.
$$
Being functions of $\,(q,\dot{q})\,$ these are not genuine equations of motion but - if not fulfilled identically - they restrict the dynamics to a subspace within the configuration-velocity space (or in geometrical terms, the tangent bundle $\,T\mathbb{Q}$). For reasons of consistency, the time derivative of these constraints must not lead outside this subspace. This condition possibly enforces further Lagrangian constraints and by this a smaller subspace of allowed dynamics, etc. 

The previous considerations are carried over to a field theory with a generic Lagrangian density ${\cal L} (Q^\alpha,\partial_{\mu}Q^\alpha)$. Rewrite the field equations as
\begin{equation}
[{\cal L}]_\alpha: =\frac{\partial {\cal L}}{\partial Q^\alpha} - \partial_\mu \frac{\partial {\cal L}}{\partial Q^\alpha_ {,\mu}} 
= \Big(\frac{\partial {\cal L}}{\partial Q^\alpha} - \frac{\partial^2 {\cal L}}{\partial Q^\alpha_{,\mu} \partial Q^\beta} Q^\beta_{, \mu} \Big) -
\frac{\partial^2 {\cal L}}{\partial Q^\alpha_{,\mu}\partial Q^\beta_{,\nu}}\,\,Q^\beta_{,\mu\nu}
:=\mathcal{V}_\alpha-\mathcal{W}_{\alpha\beta}^{\mu\nu}\,Q^\beta_{,\mu\nu}.
\label{hessFEQ} 
\end{equation} 
With the choice of the time variable $T=x^0$ the Hessian is defined by\footnote{Observe that the Hessian depends on the selection of the time variable. This indeed has the direct consequence that the number of null eigenvectors and of (primary) constraints depends on this choice.} 
\beq
{\mathcal{W}}_{\alpha\beta}:= \frac{\partial^2 {\cal L}}{\partial (\partial_0 Q^\alpha) \partial (\partial_0 Q^\beta)}.
\eeq 
If the rank of this matrix is $\,R<N\,$, it has $\,P=N$-$R\,$ null eigenvectors 
\beq 
{{\xi^\alpha}}_\rho {\mathcal{W}}_{\alpha\beta} \equiv 0. \label{nullvector}
\eeq

\subsection{Klein-Noether identities and phase-space constraints}

\subsubsection{Klein-Noether identities}

In 1918, Emmy Noether \cite{Noether:1918aa} wrote an article dealing with the consequences of symmetries of action functionals 
$$
S=\int dx^D \mathcal{L}(Q^\alpha, \partial_\mu Q^\alpha).
$$
For symmetry transformations $\bar{\delta}_S Q^\alpha$,  $\delta_S x^\mu$ her central identity is
\begin{equation}
\label{InvLocal}
[{\cal L}]_\alpha \bar{\delta}_S Q^\alpha + \partial_\mu J_S^\mu \equiv 0  
\end{equation}
with the Noether current
\begin{equation}
\label{NoeCurrent}
J_S^\mu = \frac{\partial \mathcal{L}}{\partial(\partial_\mu Q^\alpha)} \bar{\delta}_S Q^\alpha + {\cal L} \delta_S x^\mu - \Sigma^\mu_S ,
\end{equation}
where $\Sigma^\mu_S$ is a possible surface term. 

Noether's so-called second theorem deals with local symmetries, here restricted to transformations of the form\footnote{\,\,\,This form holds for our fundamental interactions as they are known today, that is for the case of Yang-Mills type theories and for general relativity. }
\begin{subequations}
\label{local}
\begin{align}
   \delta_\epsilon x^\mu & = {\cal D}^\mu_r(x) \, \epsilon^r(x) &   \\
   \delta_\epsilon Q^\alpha &= {\cal A}^\alpha_r (Q) \, \epsilon^r(x) + {\cal B}_r^{\alpha\mu}(Q)\, \epsilon^r_{,\mu}(x)
\end{align}
\end{subequations}
If one expands the Noether current int terms with the 'descriptors' $\epsilon^r$ and their derivatives, 
\begin{equation}
\label{ext_curr}
J_\epsilon^\mu = j^\mu_r \epsilon^r + k^{\mu\nu}_r \partial_\nu \epsilon^r = [j^\mu_r - \partial_\nu k^{\mu\nu}_r]\epsilon^r 
+ \partial_\nu (k^{\mu\nu}_r  \epsilon^r)
\end{equation}
Inserting this and the transformations (\ref{local}) into the invariance condition (\ref{InvLocal}), the separate vanishing of coefficients in front of the $\,\epsilon^r\,$ and those in front of their first and second derivatives gives rise to three sets of identities:
\begin{subequations}
\label{NoethKlein}
\begin{align}
k^{\mu\nu}_r + k^{\nu\mu}_r& \equiv 0. \label{Klein32} \\
[{\cal L}]_\alpha{\cal B}_r^{\alpha\mu} + j^\mu_r - \partial_\nu k^{\mu\nu}_r & \equiv 0 \label{Klein30} \\
[{\cal L}]_\alpha {\bar{\cal A}}^\alpha_r + \partial_\mu j^\mu_r & \equiv 0 \label{Klein31} ,
\end{align}
\end{subequations}
where the first two sets do not exist in the case of global symmetries. The two sets of identities (\ref{Klein31}) and (\ref{Klein30}) together imply 
\begin{equation}
\label{verallgemBianchi} 
\mathcal{N}_r = [{\cal L}]_\alpha ({\cal A}^\alpha_r - Q^\alpha_{,\mu} {\cal D}^\mu_r) - \partial_\mu ([{\cal L}]_\alpha {\cal B}^{\alpha\mu}_r)  \equiv 0.
\end{equation}
In the literature today when Noether's second theorem is mentioned, one mostly has these identities in mind.

\subsubsection{Constraints as a consequence of local symmetries}

Let us identify the terms with the highest possible derivatives of the fields $\,Q^\alpha\,$ in (\ref{verallgemBianchi}) by using the expression (\ref{hessFEQ}) which already isolates the second derivatives. A further derivative possibly originates from the last term in the Noether identity. It reads 
$\,\,{\cal B}^{\alpha\mu}_r \mathcal{W}_{\alpha\beta}^{\lambda\nu}\,Q^\beta_{,\lambda\nu\mu}.\,$ This term must vanish itself for all third derivatives of the fields, and therefore
$$
{\cal B}^{\alpha(\mu}_r \mathcal{W}_{\alpha\beta}^{\lambda\nu)}=0,
$$ 
where the symmetrization goes over $\,\mu, \lambda, \nu.$ Among these identities is the Hessian, and one finds 
\beq 
{\mathcal{B}^\alpha}_r\,{\mathcal{W}}_{\alpha\beta}=0.
\eeq 
Thus the non-vanishing $\,{\mathcal{B}}_r\,$ are null-eigenvectors of the Hessian. And comparing this with (\ref{nullvector}) there must be linear relationships
${\mathcal{B}^\alpha}_r\,=\,\lambda_r^\rho\,\,{\xi^\alpha}_\rho
$
with coefficients $\,\lambda_r^\rho\,$. In case all or some of the $\,{\mathcal{B}^\alpha}_r\,$ are zero, one can repeat the previous argumentation by singling out the terms with second derivatives, and find again that the Hessian has a vanishing determinant. Thus every action which is invariant under local symmetry transformations necessarily describes a singular system. This, however, should not lead to the impression that any singular system exhibits local symmetries: a system can become singular just by the choice of the time variable.  

\subsection{Dirac-Bergmann algorithm}

Since the fields and the canonical momenta are not independent, they cannot be taken as coordinates in a phase space as one is accustomed in the unconstrained case. This difficulty was known already by the end of the 1920's, and after unsatisfactory attempts by eminent physicists such as W. Pauli, W. Heisenberg, and E. Fermi this problem was attacked by L. Rosenfeld. As shown in the main part of this article, he undertook the very ambitious effort of obtaining the Hamiltonian version for the Einstein-Maxwell theory as a preliminary step towards quantization. But only in the late forties and early fifties did the Hamiltonian version of constrained dynamics acquire a substantially mature  form due to P. Bergmann and collaborators on the one hand \cite{Bergmann:1949aa,Bergmann:1949ab,Bergmann:1950ab,Anderson:1951aa,Bergmann:1953aa} and due to P.A.M. Dirac \cite{Dirac:1950aa,Dirac:1951aa} on the other hand.

\subsubsection{Primary constraints}

The rank of the Hessian (\ref{hessian}) being $\,R=N$-$P$\, implies that - at least locally - the equations (\ref{momentum}) can be solved for $\,R$ of the velocities in terms of the positions, some of the momenta and the remaining velocities. Furthermore, there are \,$P\,$ relations 
\begin{equation}
\label{}
\phi_\rho(q,p) = 0 \hspace{2cm} \rho=1,...,P 
\end{equation}
which restrict the dynamics to a subspace $\,\Gamma_P \subset \Gamma\,$ of the full phase space \,$\Gamma$.  These relations were dubbed \textit{primary constraints} by Anderson and Bergmann, a term suggesting that there are possibly secondary and further generations of constraints.\footnote{For many of the calculations below, one needs to set regularity conditions, namely \,(1) the Hessian of the Lagrangian has constant rank, \,(2) there are no ineffective constraints, that is constraints whose gradients vanish on $\,\Gamma_P$, \,(3) the rank of the Poisson bracket matrix of constraints remains constant in the stabilization algorithm described below. }

\subsubsection{Weak and strong equations}

It will turn out that even in the singular case, one can write the dynamical equations in terms of Poisson brackets. But one must be careful in interpreting them in the presence of constraints. In order to support this precaution, Dirac contrived the concepts of "weak" and "strong" equality.

If a function $\,F(p,q)\,$ which is defined in the neighborhood of $\,\,\Gamma_P\,$ becomes identically zero when restricted to $\,\,\Gamma_P\,$ it is called "weakly zero", denoted by $\,F \approx 0$:
$$
F(q,p) \big |_{\Gamma_P} =0 \hspace{7mm} \longleftrightarrow \hspace{7mm} F \approx 0.
$$
(Since in the course of the algorithm the constraint surface is possibly narrowed down, a better notation would be $\,F \approx_{|_{\Gamma_P}}  0\,$.) If the gradient of $\,F\,$ is also identically zero on $\,\Gamma_P\,$, $\,F\,$ is called "strongly zero", denoted by $\,F \simeq 0$:
$$
F(q,p) \big |_{\Gamma_P} =0 \hspace{5mm} \Big (\frac{\partial F}{\partial q^i}, \frac{\partial F}{\partial p_k}\Big)\big |_{\Gamma_P} = 0 \hspace{7mm} \longleftrightarrow \hspace{7mm} F \simeq 0.
$$
It can be shown that 
$$
F \approx 0 \hspace{5mm} \longleftrightarrow \hspace{5mm} F - f^\rho\phi_\rho \simeq 0.
$$
Indeed, the subspace $\,\,\Gamma_P\,$ can itself be defined by the weak equations $\phi_\rho \approx 0$.

\subsubsection{Canonical and total Hamiltonian}

Next introduce the "canonical" Hamiltonian by
$$
H_C=\,p_i\dot{q}^i- L(q,\dot{q})
$$
Its variation yields
$$
\delta H_C=\,(\delta p_i)\dot{q}^i + \,p_i\delta\dot{q}^i - \frac{\partial L}{\partial q^i}  \delta q^i  - \frac{\partial L}{\partial \dot{q}^i}  \delta \dot{q}^i = \dot{q}^i \,\delta p_i- \frac{\partial L}{\partial q^i} \, \delta q^i 
$$
(after using the definition of momenta), revealing the remarkable fact that the canonical Hamiltonian can be written in terms of $\,q$'s and $\,p$'s. No explicit dependence on any velocity variable is left, despite the fact that the Legendre transformation is non-invertible. Observe, however, that the expression for $\,\delta H_C\,$ given in terms of the variations $\,\delta q^i\,$ and $\,\delta p_i\,$ does not allow the derivation of the Hamilton equations of motion, since the variations are not independent due to the existence of primary constraints. In order that these be respected, the variation of $\,H_C\,$ needs to be performed together with Lagrange multipliers. This gives rise to define the "total" Hamiltonian 
\begin{equation}
\label{totHam}
H_T := H_C + u^\rho\phi_\rho
\end{equation}
with arbitrary multiplier functions  $\,u^\rho\,$ in front of the primary constraint functions. Varying the total Hamiltonian with respect to $(u, q, p)$ one obtains the primary constraints and
\begin{subequations}
\label{THamEqu}
\begin{align}
\label{}
    \frac{\partial H_C}{\partial p_i} + u^\rho \frac{\partial \phi_\rho}{\partial p_i}   &=  \dot{q}^i  \\
   \frac{\partial H_C}{\partial q^i} + u^\rho \frac{\partial \phi_\rho}{\partial q^i} & = - \frac{\partial L}{\partial q^i}  = - p_i
 \end{align}
 \end{subequations} 
where the last relation follows from the definition of momenta and the Euler-Lagrange equations. This recipe for treating the primary constraints with Lagrange multipliers sounds reasonable; a mathematical justification was given in Battle et al. \cite{Batlle:1986aa}. The equations (\ref{THamEqu}) are reminiscent of the Hamilton equations for regular systems. However, there are extra terms depending on the primary constraints and the multipliers. Nevertheless, (\ref{THamEqu}) can be written in terms of Poisson brackets, provided one adopts the following convention: Consider
$$
\{F, H_T\}= \{F, H_C  + u^\rho \phi_\rho\}= \{F, H_C\} + u^\rho \{F,  \phi_\rho\} + \{F, u^\rho \}\phi_\rho.
$$
Since the multipliers $\,u_\rho\,$ are not phase-space functions, the Poisson brackets $\,\{F, u^\rho \}\,$ are not defined. However, these appear multiplied with constraints and thus the last term vanishes weakly. Therefore the dynamical equations for any phase-space function $\,F(q,p)\,$ can be written as  
\begin{equation}
\label{PHam}
\dot{F}(p,q) \approx \{F, H_T \} .
\end{equation}

\subsubsection{Stability of constraints}

For consistency of a theory, one must require that the primary constraints are conserved during the dynamical evolution of the system:
\begin{equation}
\label{stab}
0 \stackrel{!}{\approx} \dot{\phi}_\rho \approx \{\phi_\rho, H_C\} + u^\sigma\{\phi_\rho,\phi_\sigma\} := h_\rho + C_{\rho\sigma}u^\sigma.
\end{equation}
There are essentially two distinct situations, depending on whether the determinant of $\,C_{\rho\sigma}\,$ vanishes (weakly) or not
\begin{itemize}
  \item $\det C \neq 0$: \,\,In this case (\ref{stab}) constitutes an inhomogeneous system of linear equations with solutions
  $u^\rho \approx -{\bar{C}}^{\rho\sigma}h_\sigma$, 
  where $\overline{C}$ is the inverse of the matrix $\,C$. Therefore, the Hamilton equations of motion (\ref{PHam}) become
  $$
 \dot{ F} \approx \{F, H_C\} - \{F, \phi_\rho\}{\bar{C}}^{\rho\sigma}\{\phi_\sigma, H_C\},
  $$
  which are free of any arbitrary multipliers.
   \item $\det C \approx 0$: \,\,In this case, the multipliers are not uniquely determined and (\ref{stab}) is only solvable if the $\,h_\rho\,$ fulfill certain relations, derived as follows: Let the rank of $\,C\,$ be $\,M$. This implies that there are $\,(P$-$M)\,$ linearly-independent null eigenvectors, i.e. $\,w^\rho_\alpha C_{\rho\sigma} \approx 0\,$ from which by (\ref{stab}) one finds the conditions
$0 \stackrel{!}{\approx} w^\rho_\alpha h_\rho.
$
These either are fulfilled or lead to a certain number $\,S'$ of new constraints 
$\phi_{\bar{\rho}} \approx 0$ \hspace{2mm} $\bar{\rho}= P+1, ..., P+S' 
$
called "secondary" constraints. The primary and secondary constraints define a hypersurface $\,\,\Gamma_2\subseteq \Gamma_P$. In a further step one has to check that the original and the newly generated constraints are conserved on $\,\Gamma_2$. This might imply another generation of constraints, defining a hypersurface $\,\Gamma_3\subseteq \Gamma_2\,$, etc., etc. In most physically relevant cases, the algorithm terminates with the secondary constraints.
\end{itemize}
The algorithm terminates when the following situation is attained: There is a hypersurface $\,\,\Gamma_C\,$ defined by the constraints
\begin{equation}
\label{}
\phi_\rho \approx_{|_{\Gamma_C}} 0 \hspace{5mm} \rho =1, ..., P  \hspace{12mm} \textrm{and} \hspace{12mm}
\phi_{\bar{\rho}} \approx_{|_{\Gamma_C}} 0 \hspace{5mm} \bar{\rho} = P+1, ..., P+S.
\end{equation}
The first set $\,\{\phi_\rho\}\,$ contains all $\,P\,$ primary constraints, the other set $\,\{\phi_{\bar{\rho}}\}\,$ comprises all secondary, tertiary, etc. constraints, assuming there are $\,S\,$ of them.  It turns out to be convenient to use a common notation for all constraints as $\,\phi_{\hat{\rho}}\,$ with $\,\hat{\rho} = 1,...,P+S$. Furthermore, for every left null-eigenvector  $\,w^{\hat{\rho}}_\alpha\,$ of the matrix ${\hat{C}}_{\hat{\rho}\rho}= \{\phi_{\hat{\rho}},\phi_\rho\}$, 
the conditions $\,\,w^{\hat{\rho}}_\alpha \{\phi_{\hat{\rho}}, H_C\}\approx_{|_{\Gamma_C}} 0\,$ are fulfilled. For the multiplier functions $\,u^\rho,$ the equations 
\begin{equation}
\label{mupl}
 \{\phi_{\bar{\rho}}, H_C\} +\{\phi_{\bar{\rho}},\phi_\rho\} u^\rho  \approx_{|_{\Gamma_C}} 0.
\end{equation}
hold. In the following, weak equality $\approx\,$ is always understood with respect to the "final" constraint hypersurface $\,\Gamma_C$.

\subsubsection{First- and second-class constraints}

Curiosity about the fate of the multiplier functions leads to the notion of first- and second-class objects. 

Some of the equations (\ref{mupl}) may be fulfilled identically, others represent conditions on the $\,u^\rho$. The details depend on the rank of the matrix $\,\hat{C}$. If the rank of $\,\hat{C}\,$ is $\,P$, all multipliers are fixed. If the rank of $\,\hat{C}\,$ is $\,K < P\,$ there are $\,P$-$K\,$ solutions of 
\begin{equation}
\label{ev}
{\hat{C}}_{\hat{\rho}\rho} V^\rho_\alpha = \{\phi_{\hat{\rho}},\phi_\rho\} V^\rho_\alpha \approx 0.
\end{equation}
The most general solution of the linear inhomogeneous equations (\ref{mupl}) is the sum of a particular solution $\,U^\rho\,$ and a linear combination of the solutions of the homogeneous part:
\begin{equation}
\label{usol}
u^\rho = U^\rho + v^\alpha V^\rho_\alpha 
\end{equation}
with arbitrary coefficients $\,v^\alpha$. Together with $\,\phi_\rho,$ also the linear combinations 
\begin{equation}
\label{FCP}
\phi_\alpha := V^\rho_\alpha \phi_\rho
\end{equation}
constitute constraint functions. According to (\ref{ev}), these have the property that their Poisson brackets with all constraints vanish on the constraint surface.

A phase-space function $\,\mathcal{F}(p,q)\,$ is said to be \textit{first class} (FC) if it has a weakly vanishing Poisson bracket with all constraints in the theory:
$$
\{\mathcal{F}(p,q),\phi_{\hat{\rho}}\} \approx 0 .
$$
If a phase-space object is not first class, it is called \textit{second class} (SC). Due to the definitions of weak and strong equality a first-class quantity obeys the strong equation
$$
\{\mathcal{F},\phi_{\hat{\rho}}\} \simeq  f^{\hat{\sigma}}_{\hat{\rho}}\,\phi_{\hat{\sigma}},
$$
from which by virtue of the Jacobi identity one infers that the Poisson bracket of two FC objects is itself an FC object

It turns out to be advantageous to reformulate the theory completely in terms of its maximal number of independent FC constraints and the remaining SC constraints. Assume that this maximal number is found after building suitable linear combinations of constraints. Call this set of FC constraints $\,\Phi_I$ ($I=1,...,L)\,$ and denote the remaining second class constraints by $\,\chi_A$. Evidence that one has found the maximal number of FC constraints is the non-vanishing determinant of the matrix built by the Poisson brackets of all second class constraints
\begin{equation}
\label{SCC}
(\Delta_{AB}) = \{\chi_A,\chi_B\}.
\end{equation} 
Rewriting the total Hamiltonian (\ref{totHam}) with the aid of  (\ref{usol}) as 
\begin{equation}
\label{finHam}
     H_T = H' +  v^\alpha \phi_\alpha \hspace{2cm} \textrm{with} \hspace{2cm} H'= H_C + U^\rho \phi_\rho,
\end{equation}
one observes that $\,H'\,$ is itself first class, and that the total Hamiltonian is a sum of a first class Hamiltonian and a linear combination of primary first class constraints (PFC). 

Consider again the system of equations (\ref{mupl}). They are identically fulfilled for the FC constraints. For a SC constraint, these equations can be written as
$$
 \{\chi_A, H_C\} + \Delta_{AB}\, u^B  \approx 0
$$
with the understanding that $\,u^B=0\,$ if $\,\chi^B\,$ is a secondary constraint (SC). For the other multipliers holds
\begin{equation}
\label{sol-u}
u^B= \overline{\Delta}^{BA}\, \{\chi_A, H_c\} \hspace{10mm} \textrm{for} \hspace{5mm} \chi_B \hspace{5mm} \textrm{primary}.
\end{equation}
where $\,\overline{\Delta}\,$ is the inverse of $\,\Delta$.  As a result, all multipliers belonging to the primary second-class constraints in $\,H'\,$ of (\ref{finHam}) are determined, and that only the $\,v^\alpha\,$ are left open: There are as many arbitrary functions in the Hamiltonian as there are (independent) primary first-class constraints (PFC).

Inserting the solutions (\ref{sol-u}) into the Hamilton equations (\ref{PHam}), they become
\begin{equation}
\label{HamConsE}
\dot{F}(p,q) \approx \{F, H_T \} \approx  \{F, H_C\}  +  \{F, \phi_\alpha \}v^\alpha -  \{F, \chi_A \}\overline{\Delta}^{AB} \{\chi_B, H_C\}.
\end{equation}

\subsection{First-class constraints and symmetries}

\subsubsection{"first-class constraints are gauge generators": perhaps some}

It was argued that a theory with local variational symmetries necessarily is described by a singular Lagrangian and that it acquires constraints in its Hamiltonian description. The previous section revealed the essential difference between regular and singular systems in that for the latter, there might remain arbitrary functions as multipliers of primary first-class constraints. An educated guess leads to suspect that these constraints are related to the local symmetries on the Lagrange level. This guess points in the right direction, but things aren't that simple. Dirac, in his famous lectures \cite{Dirac:1950aa,Dirac:1951aa} introduced an influential invariance argument by which he conjectured that also secondary first-class constraints lead to invariances. His argumentation gave rise to the widely-held view that "first-class constraints are gauge generators". Aside from the fact that Dirac did not use the term "gauge" anywhere in his lectures, later work on relating the constraints to variational symmetries revealed that a detailed investigation on the full constraint structure of the theory in question is needed; see J. Pons \cite{Pons:2005ab}.

\subsubsection{Relating Lagrangian and Hamiltonian symmetries}

Why at all should the symmetry transformations as given by (\ref{local}), that is
\begin{equation}
\label{Btrafos}
   \delta_\epsilon Q^\alpha = {\cal A}^\alpha_r (Q) \cdot \epsilon^r(x) + {\cal B}_r^{\alpha\mu}(Q)\cdot \epsilon^r_{,\mu}(x) + ...,
\end{equation}
be related to canonical transformations 
\begin{equation}
\label{consgen}
\delta_{\underline{\epsilon}} Q^\alpha = \underline{\epsilon}^I\{Q^\alpha, \Phi_I\} ?
\end{equation}
Is there a mapping between the parameter functions $\,\epsilon^r\,$ and $\,\underline{\epsilon}^I$ ? Can one specify an algorithm to calculate the generators of Noether symmetries in terms of constraints? 

Ignoring Rosenfeld, it seemed that the very first people to address these questions were Anderson and Bergmann (1951) - even before the Hamiltonian procedure for constrained systems was fully developed. N. Mukunda \cite{Mukunda:1980aa} started off from the chain (\ref{NoethKlein}) of Klein-Noether identities and built symmetry generators as linear combinations of first class primary and secondary constraints from them, assuming that no tertiary constraints are present. L. Castellani \cite{Castellani:1982aa} devised an algorithm for calculating symmetry generators for local symmetries, implicitly neglecting possible second-class constraints.

\subsection{Second-class constraints and gauge conditions}

The previous subsection dealt at length with first-class constraints because they are related to variational symmetries of the theory in question. Second-class constraints $\,\chi_A\,$ enter the Hamiltonian equations of motion (\ref{HamConsE}) without arbitrary multipliers. If there are no first-class constraints the dynamics is completely determined by
$$
\dot{F}(p,q) \approx \{F, H_T \} \approx  \{F, H_C\}   -  \{F, \chi_A \}\overline{\Delta}^{AB} \{\chi_B, H_C\}
$$
without any ambiguity.

\subsubsection{Dirac bracket}

Dirac introduced in \cite{Dirac:1950aa} a "new P.b.":
\begin{equation}
\label{defDB}
\{F, G\}^{*} := \{F, G\}  -  \{F, \chi_A \}\,\overline{\Delta}^{AB}\, \{\chi_B, G\}
\end{equation}
nowadays called the \textit{Dirac bracket} (DB). Sometimes for purposes of clarity it is judicious to indicate in the notation $\,\{F, G\}^{*}_{\Delta}\,$ that the DB is built with respect to the matrix $\,\Delta$. The Dirac bracket satisfies the same properties as the Poisson bracket, i.e. it is antisymmetric, bilinear, and it obeys the product rule and the Jacobi identity. Furthermore, the DBs involving SC and FC constraints obey
$$
\{F,\chi_A\}^{*} \equiv 0 \hspace{20mm} \{F,\Phi_I\}^{*} \approx \{F,\Phi_I\} .
$$
Thus when working with Dirac brackets, second-class constraints can be treated as strong equations. The equations of motion (\ref{HamConsE}) written in terms of DBs are 
\begin{equation}
\label{DBham}
\dot{F}(p,q) \approx \{F, H_T \}^{*}.
\end{equation}

\subsubsection{"Gauge" fixing}

The existence of unphysical symmetry transformations indicated by the presence of first-class constraints may make it necessary to impose conditions on the dynamical variables. This is specifically the case if the observables cannot be constructed explicitly - and this is notably true in general for Yang-Mills and for gravitational theories. These extra conditions are further "gauge" constraints\footnote{"gauge" is put in hyphens here since in generally covariant systems a proper name would be coordinate condition.}
\begin{equation}
\label{gc}
\Omega_a(q, p) \approx 0,
\end{equation}
where now weak equality refers to the hypersurface $\,\Gamma_R\,$ defined by the weak vanishing of all previously found first- and second-class constraints, that is the hypersurface $\,\Gamma_C\,$ together with the constraints (\ref{gc}). The idea is that the quest for stability of these constraints, namely 
$$
0 \stackrel{!}{\approx} \dot{\Omega_a} \approx  \{\Omega_a, H_C\}  +  
\{\Omega_a, \phi_\alpha \}v^\alpha -  \{\Omega_a, \chi_A \}\overline{\Delta}^{AB} \{\chi_B, H_C\}
$$
is meant to uniquely determine the multiplier $\,v^\alpha$. At least for finite-dimensional systems, the previous condition can be read as a linear system of equations which has unique solutions if the number of independent gauge constraints is the same as the number of primary FC constraints and if the gauge constraints are chosen so that the determinant of the matrix
\begin{equation}
\label{FP}
\Lambda_{\beta\alpha}: =\{\Omega_\beta, \phi_\alpha \}
\end{equation}
does not vanish\footnote{\,\,\,This determinant is related to the Faddeev-Popov determinant which plays a central r\^ole in the path integral formulation of gauge systems.}. In this case the multipliers are fixed to:
$$
v^\alpha = \overline{\Lambda}^{\alpha\gamma}
\big[- \{\Omega_\gamma, H_C\} +  \{\Omega_\gamma, \chi_A \}\overline{\Delta}^{AB} \{\chi_B, H_C\}\big].
$$

\noindent Some remarks concerning the choice of gauge constraints $\,\Omega^\alpha\,$:
\begin{itemize}
  \item The condition of a non-vanishing determinant $\,(\det (\Lambda_{\alpha\beta}) \neq 0)\,$ is only a sufficient condition for determining the arbitrary multipliers connected with the primary FC constraints.  
 
  \item The gauge constraints must not only be such that the "gauge" freedom is removed (this is guaranteed by the non-vanishing of $\,\det \Lambda$), but also the gauge constraints must be accessible: for any point in phase space with coordinates $\,(q,p),$ there must exist a transformation $\,(q,p) \rightarrow (q',p')\,$ such that $\,\Omega_\alpha(q',p') \approx 0\,$. This may be achievable only locally.
 
  \item In case of reparametrization invariance (at least one of) the gauge constraints must depend on the parameters explicitly - and not only on the phase-space variables.
  
  \item Especially in field theories it may be the case that no globally admissible (unique and accessible) gauge constraints exist. An example is given by the Gribov ambiguities, as they were first found in in Yang-Mills theories.
\end{itemize}

\bibliographystyle{plain}

\bibliography{qgrav-V19}

\end{document}